\documentclass[twocolumn,showpacs,preprintnumbers,amsmath,amssymb,nofootinbib]{revtex4}

\usepackage{graphicx}
\usepackage{dcolumn}
\usepackage{bm}

\newcommand{\re}{\text{Re}}
\newcommand{\im}{\text{Im}}
\newcommand{\qmax}{q_{\text{max}}}

\begin{document}


\title{Variational calculation of the $\bm{ppK^-}$ system based on chiral 
SU(3) dynamics}

\author{Akinobu Dot\'e}
 \email{dote@post.kek.jp}
\affiliation{%
High Energy Accelerator Research Organization (IPNS/KEK),
1-1 Ooho, Tsukuba, Ibaraki, Japan, 305-0801
}%

\author{Tetsuo Hyodo}%
 \email{thyodo@ph.tum.de}

\affiliation{
Physik-Department, Technische Universit\"at M\"unchen, 
D-85747 Garching, Germany 
}%
\affiliation{
Yukawa Institute for Theoretical Physics,
Kyoto University, Kyoto 606--8502, Japan
}%
 
 \author{Wolfram Weise}
 \email{weise@ph.tum.de}
\affiliation{
Physik-Department, Technische Universit\"at M\"unchen, 
D-85747 Garching, Germany 
}%

\date{\today}

\begin{abstract}
The $ppK^-$ system, as a prototype for possible quasibound $\bar{K}$ nuclei,
is investigated using a variational approach. Several versions of energy 
dependent effective $\bar{K}N$ interactions derived from chiral SU(3) 
dynamics are employed as input, together with a realistic $NN$ potential 
(Av18). Taking into account theoretical uncertainties in the extrapolations 
below the $\bar{K}N$ threshold, we find that the antikaonic dibaryon $ppK^-$
is {\it not} deeply bound. With the driving $s$-wave $\bar{K}N$ interaction the 
resulting total binding energy is $B(ppK^-)$ = 20 
$\pm$ 3 MeV and the mesonic decay width involving 
$\bar{K}N\rightarrow \pi Y$ is expected to be in the range 40 - 70 MeV. 
%
Properties of this quasibound $ppK^-$ system (such as density 
distributions of nucleons and antikaon) are discussed.
The $\Lambda(1405)$, as an $I=0$ quasi-bound state of $\bar{K}$ and a 
nucleon, appears to survive in the $ppK^-$ cluster. Estimates are given for 
the influence of $p$-wave $\bar{K}N$ interactions and for the width from 
two-nucleon absorption $(\bar{K}NN\rightarrow YN)$ processes. With inclusion
of these effects and dispersive corrections from absorption, the $ppK^-$ binding 
energy is expected to be in the range 20 - 40 MeV, while the total decay width 
can reach 100 MeV but with large theoretical uncertainties.
\end{abstract}

\pacs{Valid PACS appear here}
\maketitle

\section{INTRODUCTION}

In the context of low-energy QCD with $N_f = 3$ quark flavors, the study of 
possible antikaon-nuclear quasibound states is a topic of great current 
interest. Spontaneously broken chiral $SU(3)\times SU(3)$ symmetry, together
with explicit symmetry breaking by the non-zero quark masses, basically 
determines the leading couplings between the low-mass pseudoscalar meson 
octet (Nambu-Goldstone bosons in the chiral limit) and the octet of the 
ground state baryons. In particular, the Tomozawa-Weinberg chiral low-energy
theorem implies that the driving $\bar{K}N$ interaction in the isospin $I=0$
channel is  strongly attractive. Likewise, the $I=0$ $\pi\Sigma$ interaction
is attractive. The coupling between these $\bar{K}N$ and $\pi\Sigma$ 
channels is the prime feature governing the subthreshold extrapolation of 
the $\bar{K}N$ interaction. A detailed knowledge of this subthreshold 
interaction is required when exploring the possible existence of bound 
$\bar{K}$-nuclear clusters.  

The quest for strong binding and dropping antikaon masses in a nuclear 
medium has a long history which originated in early discussions of kaon 
condensation in dense matter~\cite{Kaplan:1986yq,Kaplan:1987sc} and 
continued over the years~\cite{BLRT:1994,Waas:1996fy,Waas:1997pe,
Waas:1997tw} with increasing levels of refinement~\cite{Lutz:2002}. The 
topic has recently been revived when the existence of long-lived, deeply 
bound $\bar{K}$-nuclear states was suggested using a simple potential 
model~\cite{AY_2002}. It was argued that if the $\bar{K}$-nuclear binding is
sufficiently strong to generate such systems below $\pi\Sigma$ threshold, 
their width could indeed be small~\cite{AY_2002,AMDK}. Experiments performed
in search for such states~\cite{Exp:Iwasaki,Exp:Kishimoto,Exp:Nagae} have so
far not given conclusive answers. While the search continues with more 
detailed analyses~\cite{Exp:Kishimoto2,Exp:FINUDA2,Sato:2007sb,
Exp:Iwasaki2_LamD,Exp:Iwasaki2_LamN}, ``non-exotic" final state interaction 
scenarios cannot be ruled out as interpretation of the 
data~\cite{Critical_ppK:Oset}. 

An important prototype system for these considerations is $ppK^-$, the 
simplest antikaon-nuclear cluster. It has recently been investigated using 
three-body (Faddeev) methods \cite{Faddeev:Shevchenko,Faddeev:Ikeda} and 
variational approaches \cite{ppK:Akaishi, DW-HYP06, ppK_Lstar:Oka, DHW:2008}.
Reaction studies~\cite{ppK_Reac:Koike07} have also been performed dealing 
with the actual formation of $ppK^-$. The Faddeev and variational 
calculations predict a total $ppK^-$ binding energy in a range $B\sim$ 
50 - 70 MeV, together with an estimate of the $\bar{K}NN\rightarrow\pi Y N$ 
decay width, $\Gamma \sim$ 50~-~100 MeV, depending on details of the 
interactions used.

The key issue in any such calculation is the (model dependent) extrapolation
of the $\bar{K}N$ interaction into the region far below threshold. Its 
predictive power is so far limited by the persistent lack of accurate 
constraints from data. It is in this perspective that we perform the present
variational calculation using a subthreshold effective $\bar{K}N$ 
interaction systematically derived from chiral SU(3) coupled-channel 
dynamics. 

Apart from the constraints provided by $\bar{K}N$ threshold data and 
low-energy cross sections, the only piece of information about the 
interaction below $\bar{K}N$ threshold is the $\pi\Sigma$ mass spectrum 
which is dominated by the $\Lambda(1405)$ resonance. The subthreshold 
extrapolation of the effective $\bar{K}N$ interaction has recently been 
investigated in detail~\cite{Hyodo:2007jq} from the viewpoint of chiral 
SU(3) dynamics~\cite{Kaiser:1995eg,Oset:1998it,Oller:2000fj,Lutz:2001yb,
BNW:2005}. In this approach the off-shell $\bar{K}N$ amplitude below 
threshold is governed by the strong attraction in the $\bar{K}N$ and 
$\pi\Sigma$ channels, and by the dynamical coupling of these channels. 
These couplings are again determined by the Tomozawa-Weinberg
chiral low-energy theorem. Their structure shares features with the 
early pioneering coupled-channel 
model ~\cite{Dalitz:1967fp} that used vector meson exchange interactions
(see also Ref.~\cite{Siegel:1988rq}).

Most chiral SU(3) based calculations agree
that the $\Lambda(1405)$ resonance structure, with its nominal 
position at 1405 MeV as seen in the $\pi\Sigma$ mass spectrum, is actually 
shifted to about 1420 MeV in the $\bar{K}N$ amplitude as a consequence of 
coupled-channel dynamics. This implies that the effective 
single-channel $\bar{K}N$ interaction is substantially weaker than 
anticipated in the simple phenomenological potential used previously in 
Refs.~\cite{AY_2002,ppK:Akaishi}. In those phenomenological studies, the 
local, energy-independent potential was adjusted interpreting the 
$\Lambda(1405)$ directly as a $\bar{K}N$ bound state, identifying its 
binding energy by the location of the maximum observed in the $\pi\Sigma$ 
spectrum, but ignoring strong coupled-channel effects.

In this paper we perform a variational $ppK^-$ calculation employing the new
effective $\bar{K}N$ potential derived from chiral coupled-channel 
dynamics~\cite{Hyodo:2007jq}, together with a realistic $NN$ potential.This
calculation is supposed to be complementary to the Faddeev approach with 
chiral SU(3) constraints~\cite{Faddeev:Ikeda}. The variational calculation 
gives easy access to the wave function of the bound state so that valuable 
information about the structure of the $ppK^-$ 
cluster can be extracted, whereas the elimination of the $\pi\Sigma$ channel
is required and the width of the state can only be estimated perturbatively.
The Faddeev calculation has, in turn, the advantage that the decay width of 
the quasibound state is computed consistently in the coupled-channel 
framework. Both methods therefore have their virtues and limitations which 
need to be discussed in comparison.

The present work extends and improves our previous studies~\cite{DHW:2008} 
in several directions, including further refinements in the $NN$ 
interaction, computation of density distributions, an evaluation of 
effects from $p$-wave $\bar{K}N$ interactions and an estimate of the 
$\bar{K}NN \rightarrow YN$ absorptive width. The paper is structured as 
follows. The variational framework and formalism are developed in Sec. II. 
The derivation of the effective $\bar{K}N$ interaction based on the chiral 
SU(3) coupled-channel approach is briefly summarized and discussed in 
Sec.~III. Results of our $ppK^-$ calculations are presented in Sec.~IV 
followed by a summary and conclusions in Sec.~V. 

\section{FORMALISM}

\subsection{Model wave function \label{SS:Model}}

In search for the energetically most favorable $\bar{K}NN$ configuration, 
the present variational investigation focuses on the $ppK^-$ system with 
spin and parity $J^\pi=0^-$ and isospin $(T, T_z)=(1/2, 1/2)$, where the 
parity assignment includes the intrinsic parity of the antikaon. We prepare 
the following two-component variational trial wave function: 
\begin{equation}
|\Psi \rangle = {\cal N}^{-1} [ \; |\Phi_+\rangle + C \, |\Phi_-\rangle \; ], \label{Mix}
\end{equation}
where ${\cal N}^{-1}$ is a normalization factor and $C$ is a mixing 
coefficient. The components $|\Phi_+\rangle$ and $|\Phi_-\rangle$ have the 
form 
\begin{eqnarray}
|\Phi_+ \rangle & = & 
\Phi_+ (\bm{r}_1, \bm{r}_2, \bm{r}_{\bar{K}}) 
\; \left| S_N = 0 \right\rangle \nonumber \\
& & \times \; \left| \, \left[ \, [NN]_{T_N=1} \, \bar{K} \, \right]_{T=1/2, T_z=1/2} \right\rangle , \label{Compont:+} \\
|\Phi_- \rangle & = & 
\Phi_- (\bm{r}_1, \bm{r}_2, \bm{r}_{\bar{K}}) 
\; \left| S_N = 0 \right\rangle \nonumber \\
& & \times \; \left| \, \left[ \, [NN]_{T_N=0} \, \bar{K} \, \right]_{T=1/2, T_z=1/2} \right\rangle ,  \label{Compont:-}
\end{eqnarray}
with spatial wave functions 
$\Phi_\pm (\bm{r}_1, \bm{r}_2, \bm{r}_{\bar{K}})$ multiplied by the spin 
state vector of the two nucleons and the isospin state vector of the total 
$\bar{K}NN$ system, respectively. In both components, the spin of the $NN$
pair is assumed to be zero ($S_N = 0$). The large component
$|\Phi_+ \rangle$ has isospin ($T_N=1$) of the two nucleons corresponding to
the dominant $ppK^-$ configuration, with inclusion of $pn\bar{K}^0$ through 
charge exchange. The admixture of the component $|\Phi_- \rangle$ ($T_N=0$),
with zero isospin in the $NN$ sector, can occur through a combination of 
$I=0$ and $I=1$ $\bar{K}N$ interactions and turns out to be small, typically
less than 5\%. Both components have the same total isospin $(T = 1/2)$.
 
The detailed ansatz for the spatial wave functions is chosen as follows:  
\begin{eqnarray}
\Phi_\pm (\bm{r}_1, \bm{r}_2, \bm{r}_{\bar{K}}) 
& = &   
F_N(\bm{r}_1) \, F_N(\bm{r}_2) \, F_K(\bm{r}_{\bar{K}}) 
G(\bm{r}_1, \bm{r}_2) \nonumber \\
& & \times \left[ 
\; H_1 \, (\bm{r}_1, \bm{r}_{\bar{K}}) \, H_2 \, (\bm{r}_2, \bm{r}_{\bar{K}}) \right. \nonumber \\
& & \left. \; \; \; \; \pm
\; H_2 \, (\bm{r}_1, \bm{r}_{\bar{K}}) \, H_1 \, (\bm{r}_2, \bm{r}_{\bar{K}})
\; \right] ~.\label{SpatialWF}
\end{eqnarray}
Here $F_N(\bm{r}_i)$ ($i=1,2$) and $F_K(\bm{r}_{\bar{K}})$ are trial 
functions  describing the localization of the nucleons and the kaon, 
respectively. Their forms are assumed to be single Gaussians:  
\begin{equation}
F_N(\bm{r}_i)= \exp[-\mu \, \bm{r}^2_i]~, \; \;  
F_K(\bm{r}_{\bar{K}}) =  \exp[-\gamma \, \bm{r}^2_{\bar{K}}] ~. \label{SPmotion}
\end{equation}
The $\mu$ and $\gamma$ in Eq.~(\ref{SPmotion}) are treated as independent 
parameters. Correlations between the two nucleons, and between the $\bar{K}$
and each nucleon, are described by corresponding correlation functions: 
\begin{eqnarray}
G(\bm{r}_1, \bm{r}_2) & = & 
1 - \sum_{n=1}^{N_N} 
g_n \exp\left[-\lambda_n \left(\bm{r}_1 - \bm{r}_2
\right)^2 \right]  \label{NNcorr} \\
H_a(\bm{r}_i, \bm{r}_{\bar{K}}) & = & 
1 + \sum_{n=1}^{N_K} 
h_{a,n} \exp\left[-\nu_n \left(\bm{r}_i - \bm{r}_{\bar{K}}
\right)^2 \right]  \label{KNcorr}
\end{eqnarray}
The $NN$ correlation function, $G(\bm{r}_1, \bm{r}_2)$, is prepared to 
account for the strong short-distance repulsion in the $NN$ interaction 
which keeps the two nucleons apart. The $\bar{K}N$ correlation functions, 
$H_a(\bm{r}_i, \bm{r}_{\bar{K}})$, are given the flexibility to adjust 
themselves appropriately to the attractive antikaon-nucleon interaction. 

One notes that the spatial wave functions
$\Phi_\pm (\bm{r}_1, \bm{r}_2, \bm{r}_{\bar{K}})$, Eq.~(\ref{SpatialWF}), 
are even or odd under exchange of the two nucleons: 
\begin{equation}
\Phi_\pm (\bm{r}_2, \bm{r}_1, \bm{r}_{\bar{K}}) 
= \, \pm \, \Phi_\pm (\bm{r}_1, \bm{r}_2, \bm{r}_{\bar{K}}) ~.
\end{equation}
The $NN$ pair in $|\Phi_+\rangle$ ($|\Phi_-\rangle$) is thus in a 
singlet-even (singlet-odd) state. The spatial wave function is rotationally 
symmetric, i.e. the total orbital angular momentum is $L = 0$.

The trial wave function has the following real-valued variational 
parameters: $C$ in Eq.(\ref{Mix}), $\mu$ and  $\gamma$ in 
Eq.~(\ref{SPmotion}), $\{ g_n, \lambda_n \}$ ($n=1, ... , N_N$) in
Eq.~(\ref{NNcorr}), and $\{ h_{a,n}, \nu_n \}$ ($n=1, ... , N_K$) in 
Eq.~(\ref{KNcorr}). The range parameters of the Gaussians in the $NN$ and 
$\bar{K}N$ correlation functions are organized as 
\begin{equation}
\lambda_n = \lambda_1 \cdot 
\left( \frac{\lambda_{N_N}}{\lambda_1} \right) 
^ {\frac{n-1}{N_{N}-1}},
\; \; ~~~
\nu_n = \nu_1 \cdot 
\left( \frac{\nu_{N_K}}{\nu_1} \right) 
^ {\frac{n-1}{N_{K}-1}},
\end{equation} 
independently for each of the $NN$ or $\bar{K}N$ sectors. Various 
combinations of parameter sets for the Gaussian ranges, $\mu$, $\gamma$, 
($\lambda_1$, $\lambda_{N_N}$) and ($\nu_1$, $\nu_{N_K}$) have been tried. 
For each combination of the range parameters, we find a set of parameters 
$C$ and $\{ g_n; h_{a,n}\}$ which minimizes the expectation value of
the total Hamiltonian, using the Simplex method~\cite{Simplex}. This 
determines the $ppK^-$ bound state of minimal energy, if existent. 

A remark should be added concerning the treatment of the center-of-mass (CM) motion.  
Given the independence of the variational parameters $\mu$ and $\gamma$ in 
Eq.~(\ref{SPmotion}), the CM wave function cannot simply be separated by 
factorization in the present model. A complete separation is possible only 
in the special case $\gamma / \mu = m_K / M_N$, where $M_N$ ($m_K$) is the 
nucleon (kaon) mass. In the actual calculations we have confirmed that this
relation between $\gamma$ and $\mu$ turns out to be satisfied quite 
accurately even though it has not been imposed from the beginning. The 
variational procedure favors indeed a wave function in which the CM motion 
factorizes as it should. One typically finds $\mu\simeq0.2$ fm$^{-2}$ and 
$\gamma\simeq0.1$ fm$^{-2}$, using the $\bar{K}N$ interaction based on 
chiral SU(3) dynamics and described in Sec. III.
 
\subsection{Hamiltonian}

The Hamiltonian used in the present study is of the form 
\begin{equation}
\hat{H} = \hat{T} + \hat{V}_{NN} + {\rm Re}\,\hat{V}_{\bar{K}N} - \hat{T}_{CM}~.  
\end{equation}
Here $\hat{T}$ is the total kinetic energy: 
\begin{equation} 
\hat{T}=\frac{\hat{\bm{p}}^2_1+ \hat{\bm{p}}^2_2}{2M_N} + 
\frac{\hat{\bm{p}}^2_{\bar{K}}}{2m_K}~. 
\end{equation} 
The energy of the center-of-mass motion,
\begin{equation} 
\hat{T}_{CM}=\frac{\left( \hat{\bm{p}}_1 + \hat{\bm{p}}_2 + 
\hat{\bm{p}}_{\bar{K}} \right)^2}{2 \, \left( 2M_N+m_K \right)}~,
\end{equation}
is subtracted.

As a realistic nucleon-nucleon interaction $\hat{V}_{NN}$ we choose the 
Argonne v18 potential (Av18)~\cite{Av18}. Since the total spin of the two 
nucleons is restricted to zero as explained in the previous section, the 
tensor, $LS$ and $(LS)^2$ potentials do not contribute. We thus employ the 
central, $L^2$ and spin-spin parts of the Av18 potential: 
\begin{align}
\hat{V}_{NN}   = &  \sum_{X=^1E, ^1O} \hat{P}(X) \nonumber \\  
 & \times \left[ \; v^c_{X}(r) \, + v^{L2}_{X}(r) \, \hat{\bm{L}}^2 \, 
 + v^{SS}_{X}(r) \, \hat{\bm{\sigma}}_1 \cdot \hat{\bm{\sigma}}_2 \right], 
\end{align}
where $\hat{P}(X)$ is a projection operator onto the singlet-even ($^1E$) or
singlet-odd ($^1O$) state, and $\hat{\bm{L}}$ is the orbital angular momentum 
operator for the relative coordinate between two nucleons. The central and 
$L^2$ potentials [$v^c_{X}(r)$ and $v^{L2}_{X}(r)$] are identified with the 
phenomenological short- and intermediate-range parts of the Av18 potential, 
Eq.~(20) in Ref.~\cite{Av18}. The long-range spin-spin term comes from 
one-pion-exchange, Eq.~(17) in Ref.~\cite{Av18}. We ignore the 
electromagnetic part of Av18. Both $^1E$ and $^1O$ potentials are taken 
into account since our model wave function includes both types of $NN$ 
states. In the present study the $r$-dependence of each of the potential 
terms is well fitted by a series of Gaussians (see Appendix~\ref{App:NN}).
The most pronounced feature is the strong short-distance repulsion in the 
singlet-even central potential.  

The energy-dependent effective $s$-wave $\bar{K}N$ interaction 
$\hat{V}_{\bar{K}N}$ is represented as
\begin{align}
\hat{V}_{\bar{K}N}  =&~ \hat{v}(\bar{K}N_1) +\hat{v}(\bar{K}N_2)~, \\
\hat{v}(\bar{K}N)  =& \sum_{I=0,1} \hat{P}_I (\bar{K}N)  \nonumber \\
&\times~ v_{\bar{K}N}^I (\sqrt{s}) \, \exp\left[-(\bm{r}_{\bar{K} N} / a_s)^2 \right], 
\label{eq:KbarNint}
\end{align} 
where $\bm{r}_{\bar{K} N} = \bm{r}_{\bar{K}}-\bm{r}_N$ for each of the two 
nucleons, $N_1$ and $N_2$, and $\hat{P}_I (\bar{K}N)$ is the isospin 
projection operator for the $\bar{K}N$ pair. In the present work, the radial
dependence of the $\bar{K}N$ potential is assumed to be a single Gaussian 
form with range parameter $a_s$. The interaction strength 
$v_{\bar{K}N}^I (\sqrt{s})$ is a function of the (off-shell) center-of-mass 
energy variable  $\sqrt{s}$ of the $\bar{K}N$ system. This interaction, 
extrapolated into the subthreshold region, is a key issue in the present 
paper and will be specified in greater detail in a separate 
Sec.~\ref{sec:KbarNpotential}.

\subsection{Calculational procedure}

The energy dependence of the $\bar{K}N$ interaction requires a 
self-consistent variational procedure to minimize the energy of the 
$\bar{K}NN$ system. This is done in the same way as in our previous 
work~\cite{DW-HYP06}.

We introduce an auxiliary (non-observable) antikaon ``binding energy" $B_K$ 
to control the CM energy $\sqrt{s}$  of the $\bar{K}N$ subsystem within the 
$ppK^-$ cluster. This $B_K$ is defined as 
\begin{equation}
- B_K \equiv \langle \Psi | \hat{H} | \Psi \rangle 
- \langle \Psi | \hat{H}_{N} | \Psi \rangle~,
\label{eq:BK}
\end{equation}
where $\hat{H}_{N}$ is the nucleonic part of the Hamiltonian, 
\begin{eqnarray}
\hat{H}_{N} & = & \hat{T}_{N} + \hat{V}_{NN} - \hat{T}_{CM, N}, \\
\hat{T}_{N} & = & \frac{\hat{\bm{p}}^2_1 + \hat{\bm{p}}^2_2}{2M_N}, \; \;  
\hat{T}_{CM, N}=\frac{\left( \hat{\bm{p}}_1 + \hat{\bm{p}}_2 \right)^2}{4M_N}~.  
\end{eqnarray}
The relation between the $\bar{K}N$ two-body energy $\sqrt{s}$ and $B_K$ 
within the three-body system is not $a$ $priori$ fixed. In general,
\begin{equation}
    \sqrt{s} = M_N + m_K - \eta\, B_K  \label{DEF}~~,
\end{equation}
where $\eta$ is a parameter describing the balance of the antikaon energy 
between the two nucleons of the $\bar{K}NN$ three-body system. One expects 
$1/2 \le\eta \le1$. The upper limit ($\eta = 1$) corresponds to the case in 
which the antikaon field collectively surrounds the two nucleons, a 
situation encountered in the limit of static (infinitely heavy) nucleon 
sources. In the lower limit ($\eta = 1/2$) the antikaon energy is split 
symmetrically half-and-half between the two nucleons. We investigate both 
cases and label them ``Type I" and ``Type II", respectively:
\begin{eqnarray}
    {\rm Type \; I \; :} ~~~~~~~~\sqrt{s}& = &M_N + m_K - B_K~~, 
    \label{DEF1}\\
    {\rm Type \; II \; :} ~~~~~~~~\sqrt{s}& = &M_N + m_K - B_K/2~~. 
    \label{DEF2}
\end{eqnarray}

The calculations then proceed as follows. First, assume $B_K$ to be some 
trial starting value, $B_K^{(0)}$. Given a relation between $B_K$ and 
$\sqrt{s}$, the strength of the $\bar{K}N$ potential is now fixed, and the 
Hamiltonian is determined. Then the variational calculation is performed to 
find the state of minimal energy. Given that state, a new antikaon binding 
energy, $B_K^{(1)}$, is evaluated with the wave function so obtained. Then 
one examines whether $B_K^{(1)}$ coincides with $B_K^{(0)}$. If not, a 
different starting value $B_K^{(0)}$ is chosen and the procedure is repeated
until $B_K^{(1)} = B_K^{(0)}$ is satisfied at an acceptable level of 
accuracy. 

The $\bar{K}N$ potential is in general complex. In order to perform the 
variational calculation of the energy and the ``bound state" $|\Psi\rangle$,
the real part of the potential,  ${\rm Re} \,\hat{V}_{\bar{K}N}$, is used as
a starting point. The decay width $\Gamma$ of that state is then calculated 
perturbatively by taking the expectation value of the imaginary part of the 
$\bar{K}N$ potential:
\begin{equation} 
\Gamma = -2 \; \langle \Psi | \, {\rm Im}\,\hat{V}_{\bar{K}N} \, | \Psi \rangle~. 
\label{width1}
\end{equation}
At this stage the width $\Gamma$ represents the mesonic two-body decay 
channels ($\bar{K}N \rightarrow \pi\Sigma, \, \pi\Lambda$) within the 
$\bar{K}NN$ three-body system. The non-mesonic absorption width for 
$\bar{K}NN \rightarrow \Sigma N, \, \Lambda N$ will be treated separately.

\section{EFFECTIVE $\bar{\bm{K}}\bm{N}$ INTERACTION}
\label{sec:KbarNpotential}

\subsection{$\bar{\bm{K}}\bm{N}$ potential based on chiral SU(3) dynamics}

Here we discuss the effective $\bar{K}N$ potential developed in
Ref.~\cite{Hyodo:2007jq}. This potential has been systematically constructed 
using chiral SU(3) coupled-channel calculations which successfully describe 
$S=-1$ meson-baryon scattering and the properties of the dynamically 
generated $\Lambda(1405)$ resonance. The formulation of the coupled-channel 
approach is briefly sketched in Appendix~\ref{App:ccSU(3)}. Starting from 
this coupled-channel framework, a complex and energy-dependent interaction 
kernel $V^{\text{eff}}$ is derived in the single $\bar{K}N$ channel, such 
that the full coupled-channel $s$-wave $\bar{K}N$ scattering amplitude is 
exactly reproduced by solving the single-channel equation with 
$V^{\text{eff}}$. Then this effective interaction kernel is approximated by 
an equivalent local $\bar{K}N$ potential used in the Schr\"odinger equation 
(with reduced $\bar{K}N$ mass $\mu$),
\begin{equation}
    -\frac{d^2 u(r)}{dr^2}
    + 2\mu\left[U(r,\sqrt{s}) + B\right] u(r) = 0~  .
    \label{eq:Schroedinger}
\end{equation}
The potential $U(r,\sqrt{s})$ is expressed by the effective interaction 
kernel $V^{\text{eff}}(\sqrt{s})$ together with a normalized spatial 
distribution $g(r)$,
\begin{equation}
    U(r,\sqrt{s}) = \frac{g(r)}{2\,\tilde{\omega}}\frac{M_N}{\sqrt{s}} 
    \,V^{\text{eff}}(\sqrt{s}) ,
    \label{eq:KbarNpot0}
\end{equation}
where $r$ is the relative coordinate of the $\bar{K}N$ system and
$\tilde{\omega} = [s^2 - (M_N^2 - m_K^2)^2]/(4s^{3/2})$ is the reduced 
energy of the $\bar{K}N$ two-body system.

An important observation in Ref.~\cite{Hyodo:2007jq} and earlier work is 
that the $\pi\Sigma$ diagonal coupling is strong enough to generate a 
resonance in the single (elastic) $\pi\Sigma$ channel. In the strongly 
interacting $\bar{K}N \leftrightarrow \pi\Sigma$ system, the $\Lambda(1405)$
appears as a $\bar{K}N$ bound state embedded in the resonant $\pi\Sigma$ 
continuum. The experimentally observed broad spectrum in the $\pi\Sigma$ 
channel, with its maximum at 1405 MeV, is dominated by the inherent 
$\pi\Sigma$ interaction, while the resonant structure in the $\bar{K}N$ 
amplitude, governed by the inherent $\bar{K}N$ interaction, actually appears
around $1420$ MeV. The $\bar{K}N$ ``binding energy'' commonly associated 
with the $\Lambda(1405)$ is therefore not 27 MeV but only less than half of 
this naive estimate. This small $\bar{K}N$ binding energy implies a weaker 
attractive potential, only about half as strong as the phenomenological 
potential of Refs.~\cite{AY_2002,ppK:Akaishi}. One should nonetheless note 
that both phenomenological and chiral potentials reproduce the existing 
experimental data around threshold. Their qualitatively different 
subthreshold extrapolations result from the fact that the coupled-channels
framework, constrained by the chiral effective Lagrangian, induces off-shell
dynamics which is very different from the purely phenomenological approach.

Here we choose the spatial distribution of the potential as a Gaussian form
\begin{equation}
    U(r,\sqrt{s}) 
    = v^{I}_{\bar{K}N}(\sqrt{s})
    \, \exp\left[-(r / a_s)^2 \right] , 
    \label{eq:KbarNpot1}
\end{equation}
where $a_s$ is the range parameter. The strength of the approximate local 
potential is related to the interaction kernel $V^{\text{eff}}(\sqrt{s})$ 
defined in Ref.~\cite{Hyodo:2007jq} as
\begin{equation}
    v^{I}_{\bar{K}N}(\sqrt{s})
    =  \frac{M_N}
    {2\pi^{3/2}\,a_s^3 \,\tilde{\omega}\sqrt{s}}\,V^{\text{eff}}(\sqrt{s})~.
    \label{eq:uncorrected} 
\end{equation}
The potential so obtained is complex and energy dependent, 
reflecting the elimination of the other (mainly $\pi\Sigma$) channels. 
The center-of-mass energy $\sqrt{s}$ is related to the 
binding energy $B$ of the $\bar{K}N$ two-body system $B$ as
\begin{equation}
    B = -\sqrt{s}+M_N+m_K~ .
    \nonumber
\end{equation}
We choose the range parameter $a_s$ such that the resonance structure is 
reproduced at the position predicted by the full chiral dynamics 
calculation. The scattering amplitudes for both $I=0$ and $I=1$ around 
$\bar{K}N$ threshold are also well reproduced by this potential. 

\begin{figure}[tbp]
    \centering
    \includegraphics[width=0.5\textwidth,clip]{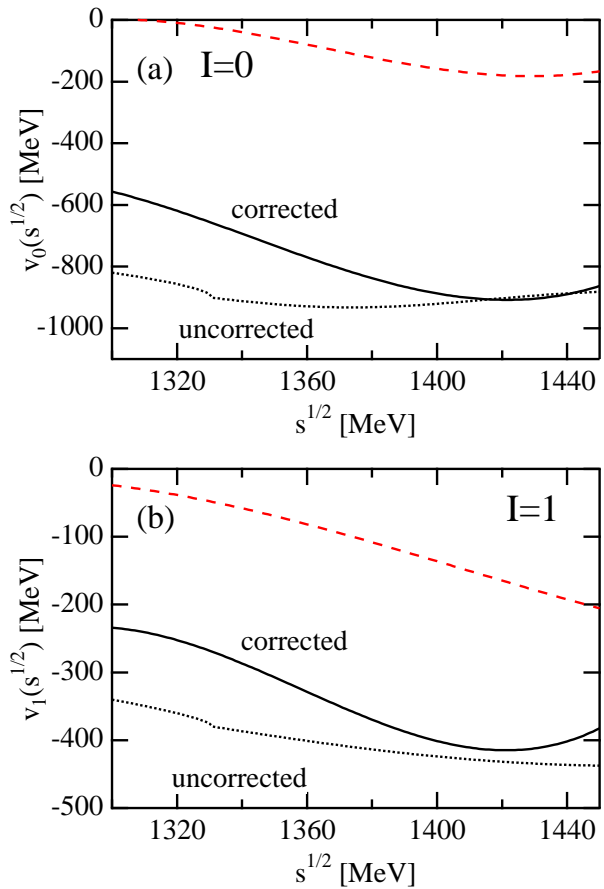}
    \caption{\label{fig:KbarNpotential}
    (Color online) Strength of the ``corrected'' (solid lines)
    and ``uncorrected'' (dotted lines)
    potentials at $r=0$ with HNJH model \cite{Hyodo:2002pk}. 
    The real parts are shown as solid/dotted
    lines. Imaginary parts are given as dashed lines. 
    (a): $I=0$ channel, (b): $I=1$ channel.}
\end{figure}

However, the effective single-channel $\bar{K}N$ effective interaction 
$V^{\text{eff}}$ is generally {\it non-local}. Naive translation of 
$V^{\text{eff}}$ into an approximate {\it local} form does not guarantee 
that this local potential (we refer to it as ``uncorrected") reproduces the 
$\bar{K}N$ amplitude of the full coupled-channel calculation over a wide 
range of subthreshold energies. It is indeed found that simple extrapolation
of the local potential~\eqref{eq:uncorrected} to the deep subthreshold 
region, $\sqrt{s}<1400$ MeV, significantly overestimates the scattering 
amplitude in comparison with that of the original coupled-channel 
approach~\cite{Hyodo:2007jq}. Compensation of this deficiency requires 
modifying the strength of the real part of the potential and introducing 
extra energy dependence. The strengths of these ``corrected'' potentials in 
the $I =0$ and $I = 1$ channels are parametrized by  polynomials as
\begin{equation}
    v^I_{\bar{K}N}(\sqrt{s})
    = K_{I,0}+ K_{I,1}\ s^{1/2} + K_{I,2}\ s+ K_{I,3}\ s^{3/2}~. 
    \label{eq:corrected}
\end{equation}
The coefficients $K_{I,i}$ are given in Ref.~\cite{Hyodo:2007jq}. The 
strengths of the ``corrected'' and ``uncorrected'' potentials at $r=0$ are 
shown in Fig.~\ref{fig:KbarNpotential}, based on the chiral model of
Ref.~\cite{Hyodo:2002pk}.  For the subsequent variational three-body 
calculation we always use the ``corrected" potentials.

In order to estimate theoretical uncertainties we use altogether four 
different variants of chiral dynamics calculations from which we derive 
equivalent energy-dependent local potentials, 
ORB~\cite{Oset:2001cn}, HNJH~\cite{Hyodo:2002pk}, 
BNW~\cite{Borasoy:2005ie}, and BMN~\cite{Borasoy:2006sr}. All these models 
reproduce the total cross sections for elastic and inelastic $K^-p$ 
scattering, threshold branching ratios, and the $\pi\Sigma$ mass spectrum 
associated with the $\Lambda(1405)$. Differences among those models mainly 
stem from the lack of accurate data for the $\pi\Sigma$ spectrum and from 
differences in details of the fitting procedures. The values of the range 
parameters for these models are shown in the second row in 
Table~\ref{tbl:KbarNresult}. Additional uncertainties, concerning higher 
order terms in the interaction kernel derived from chiral SU(3) dynamics, 
are estimated to be about 20 \% around $\sqrt{s}\sim 1360$ MeV, based on the
systematic study of such higher order corrections in 
Ref.~\cite{Borasoy:2005ie}. 

\subsection{Structure of the $\Lambda(1405)$}

\begin{table}[t]
    \centering
    \caption{Range parameters $a_s$ of the local $\bar{K}N$ potential, 
    the self-consistent $\bar{K}N$ binding energy $B$, and the root
     mean distance $\sqrt{\langle r^2 \rangle}$ between 
     antikaon and nucleon. 
     The energies $B_1$ and $B_2$ are determined, respectively, 
     by the zero of the real part and the maximum of the imaginary 
     part of the full amplitude with complex potential (see text).
    \label{Lambda1405}}    
    \begin{ruledtabular}
        \begin{tabular}{lrrrr}
         & ORB~\cite{Oset:2001cn} 
	 & HNJH~\cite{Hyodo:2002pk} 
	 & BNW~\cite{Borasoy:2005ie} 
	 & BMN~\cite{Borasoy:2006sr} \\
        \hline\\
        $a_s$ [fm]  & 0.52 & 0.47 & 0.51 & 0.41 \\
        $B$ [MeV]  &  $11.8$ & $11.5$ & $9.97$ & $13.3$ \\
        $\sqrt{\langle r^2 \rangle}$ [fm]  
	& $1.87$ & $1.86$ & $1.99$  & $1.72$   \\
	\\
	\hline\\
        $B_1$ [MeV]  &  $16.5$ & $15.5$ & $17.0$ & $16.7$ \\
        $B_2$ [MeV]  &  $17.8$ & $16.9$ & $19.8$ & $18.9$ \\
        \\
    \end{tabular}
    \end{ruledtabular}
    \label{tbl:KbarNresult}
\end{table}

Let us now examine the two-body $\bar{K}N$ system with 
$I=0$ where the $\Lambda(1405)$ is generated dynamically below the 
$\bar{K}N$ threshold. In order to study the structure of the 
$\Lambda(1405)$, we first solve the Schr\"odinger 
equation~\eqref{eq:Schroedinger} with the real part of the potential,  
$\re [U(r,\sqrt{s})]$. This treatment appears to be justified by the 
relatively small imaginary part of the potential as seen in 
Fig.~\ref{fig:KbarNpotential}. Corrections from the dispersive shift of the 
binding energy induced by the imaginary part of the potential will be 
estimated as we move along.

The self-consistent results for the binding energies are summarized in 
Table~\ref{tbl:KbarNresult} together with the root mean distance 
($\sqrt{\langle r^2 \rangle}$) between antikaon and nucleon which form the
$\Lambda(1405)$. These results are produced with the ``corrected'' 
potentials, but the ``uncorrected'' ones give essentially the same output 
since the difference in strength between the ``corrected'' and ``uncorrected'' 
potentials is small in the energy region relevant to the $\bar{K}N$ bound
state ($\sqrt{s}\sim 1420$ MeV). The typical $\bar{K}N$ binding energy and 
root mean distance found with the chiral potential,
\begin{equation}
    B \sim 12 \text{ MeV} ~,
    \quad \sqrt{\langle r^2\rangle}\sim 1.9 \text{ fm} ~,
    \label{eq:realpart}
\end{equation}
should be compared with the results of the phenomenological 
model~\cite{ppK:Akaishi}: $B = 27$ MeV and $\sqrt{\langle r^2\rangle}= 1.36$
fm. The small binding obtained with the present potential is related to the
strong $\pi\Sigma$ interaction in chiral dynamics, as discussed in 
Ref.~\cite{Hyodo:2007jq}. Since the binding energy is smaller, the size of 
the bound state becomes correspondingly larger than that of the 
phenomenological model. 

It is instructive to recall the study of electromagnetic properties of the 
$\Lambda(1405)$ in chiral dynamics~\cite{Sekihara:2008ap}, where a 
relatively large electric mean squared radius has been reported. Although 
the electric mean squared radius is not directly comparable to the mean 
distance of antikaon and nucleon, qualitative agreement of these independent
size estimates gives support to the chiral effective potential introduced here, while the radius 
in Ref.~\cite{Sekihara:2008ap} was computed with the full chiral coupled-channel
amplitudes.

\subsection{Dispersive effects induced by the imaginary part of the 
potential}\label{subsec:dispersive}

The variational calculation of the $ppK^-$ three-body problem has the 
disadvantage (unlike the Faddeev approach) that only the real part of the 
complex $\bar{K}N$ potential can be handled as input whereas the imaginary 
part must be treated perturbatively. It is therefore mandatory to estimate 
the systematic uncertainties caused by this limitation.

In the $\bar{K}N$ two-body case, the scattering amplitude can easily be 
obtained by solving Eq.~\eqref{eq:Schroedinger} with the full complex potential. 
Effects of the dispersive shift on the binding energy, induced by the 
imaginary part of this potential, can then be examined by comparing the 
full result (with complex potential) to the one obtained using only the real
part of the potential, see Eq.\eqref{eq:realpart}. The energy, $B_1$, of the 
subthreshold $\bar{K}N$ state generated by the complex potential can be 
deduced from the zero of the real part of the $\bar{K}N$ scattering amplitude. For 
comparison we also check the position of the maximum, $B_2$, of the 
imaginary part of that amplitude, which coincides with $B_1$ if the 
non-resonant background is small.

The calculated values of $B_1$ (and $B_2$) are listed, for all models 
considered, in Table~\ref{tbl:KbarNresult}. Comparing $B_1$ with the binding
energies $B$ found with the real parts of the potentials, we conclude that 
the dispersive effects may increase the binding energy of the $\bar{K}N$ 
two-body quasibound state by
\begin{equation}
    \Delta B \sim 3~\text{-}~7 \text{ MeV}
    \label{eq:dispersive} ,
\end{equation}
or $\Delta B\sim 6$ - 10 MeV when comparing $B_2$ and $B$. We keep this 
shift $\Delta B$ in mind for a discussion of systematic 
uncertainties when we now turn to the $\bar{K}NN$ three-body system.

\section{RESULTS \label{RESULT}}

This section presents the results of our variational calculations of the 
quasibound $ppK^-$ system. Four variants of effective $\bar{K}N$ 
interactions have been employed: ORB, HNJH, BNW and BMN as explained in the 
previous section. These interactions are translated into local, 
energy-dependent potentials. In all cases we have used the ``corrected'' 
versions of the potentials which properly reproduce the $\bar{K}N$ 
scattering amplitude computed with full chiral coupled-channel dynamics. In 
the self-consistent treatment of the energy dependence of the interaction, 
both options, ``Type I'' [Eq.~(\ref{DEF1})] and ``Type II'' 
[Eq.~(\ref{DEF2})] for the relationship between $\sqrt{s}$ and $B_K$ have 
been tried. With these different choices for on-shell equivalent potentials 
and different options for handling off-shell energies within the three-body 
system, we are in a position to give rough estimates of the uncertainties 
associated with the required subthreshold extrapolations.
In all calculations, the convergence of the Gaussian expansions 
(\ref{NNcorr}) and (\ref{KNcorr}) has been checked and found satisfactory 
with $N_N=N_K=5$.

\subsection{Total binding energy and decay width}

\begin{table}[t]
\caption{\label{tab:Res/Summary} 
Results of variational calculations for $ppK^-$ binding energies and 
($\bar{K}NN\rightarrow \pi Y N$) decay widths $\Gamma$ based on four 
versions of chiral effective $\bar{K}N$ potentials. ``ORB'' etc. indicate 
the chiral SU(3) model used in each case. The ``corrected'' equivalent local
potentials have been adopted throughout, with ranges $a_s$ given in 
Table~\ref{tbl:KbarNresult}. ``Type I (II)'' refers to the ansatz for the 
relationship between $\sqrt{s}$ and $B_K$ (see text).
}
\begin{ruledtabular}
\begin{tabular}{lrrrr}
        &ORB & HNJH & BNW & BMN \\  
\hline
          &   & &    &    \\
``Type I''&   & &    &    \\
$B(ppK^-)$ [MeV] & 18.9 & 16.9 & 18.1 & 16.6 \\
$\Gamma$ \hspace{1.05cm} [MeV] & 53.1 & 47.0 & 60.4 & 38.8 \\ \\
\hline
            &   & &    &    \\
``Type II'' &   & &    &    \\
$B(ppK^-)$ [MeV] & 22.7 & 20.8 & 20.7 & 21.7 \\
$\Gamma$ \hspace{1.05cm} [MeV] & 64.2 & 58.3 & 71.4 & 53.1 \\
\\
\end{tabular}
\end{ruledtabular}
\end{table}

\begin{figure}[t]
\includegraphics[width=8cm,angle=0]{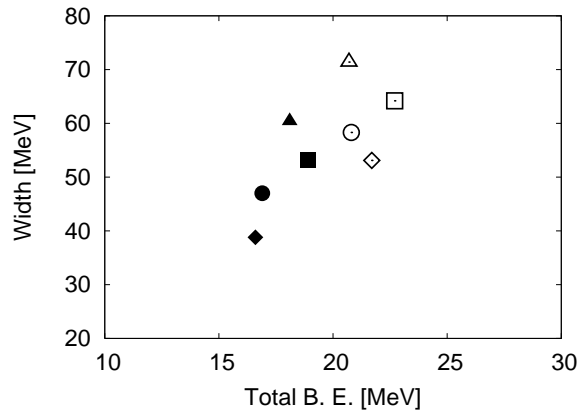}
\caption{\label{fig:BEvsW} Distribution of total binding energy and mesonic 
decay width. The models ``ORB'', ``HNJH'', ``BNW'' and ``BMN'' are shown 
with symbols square, circle, triangle and diamond, respectively. Closed 
(Open) symbols indicate the ``Type I'' (``Type II'') ansatz.}  
\end{figure}

The results of self-consistent solutions for total $ppK^-$ binding energies 
are summarized together with the $\bar{K}NN \rightarrow \pi Y N$ decay widths in 
Table~\ref{tab:Res/Summary} and depicted in Fig.~\ref{fig:BEvsW}. In all 
present calculations the binding energies turn out to be quite modest. 
Deeply bound, narrow  $ppK^-$ states are not seen to develop. The variational calculations,
with $\re V_{\bar{K}N}$ as previously specified and $\im V_{\bar{K}N}$ treated according to 
Eq.(\ref{width1}), predicts total binding energies and decay widths in the range
\begin{equation}
B(ppK^-) \sim 20\pm 3~\text{MeV}~~\text{and} 
~~\Gamma \sim 40~\text{-}~70~\text{MeV} \nonumber
\end{equation}
for all cases studied. The ``Type II'' ansatz favors slightly stronger 
binding (20 - 23 MeV) than the ``Type I'' option (16 - 19 MeV). Different 
potentials produce binding energy variations within only about two MeV for 
each given ``Type I''  or ``Type II'' set. 

It is instructive to examine the detailed decomposition of the total $ppK^-$
energy into kinetic and potential energies of the antikaon and two-nucleon 
subcomponents. These sets of numbers are given in 
Tables~\ref{tab:Res/Smooth/Wf1} and \ref{tab:Res/Smooth/Wf2} for the four 
variants of chiral SU(3) based models mentioned previously. Shown 
(separately for the ``Type I'' and ``Type II'' options) are the total 
kinetic energy, $E_{kin} =  \langle\Psi|\hat{T} - \hat{T}_{CM}|\Psi\rangle$,
the nuclear part of the kinetic energy, 
$T_{nuc} =  \langle\Psi|\hat{T}_N - \hat{T}_{CM,N}|\Psi\rangle$, the 
antikaon binding energy $B_K$ defined in Eq.~(\ref{eq:BK}), the expectation 
value of the $\bar{K}$-nuclear potential energy, 
$V(\bar{K}N) = \langle\Psi|\text{Re}\,\hat{V}_{\bar{K}N}|\Psi\rangle$, and 
the contribution from the nucleon-nucleon interaction, 
$V(NN) = \langle\Psi|\hat{V}_{NN}|\Psi\rangle$.

Several interesting observations can be made. First, the kaon 
``binding energy" $B_K$ is in the range 40 - 50 MeV for all cases studied. 
Note, however, once again that $B_K$ is {\it not} an observable in the 
$\bar{K}NN$ three-body system. In fact the observable total $ppK^-$ binding 
energy is less than half of $B_K$. The nucleons are the slow movers in the 
quasibound compound: their kinetic energies add up to only about 40 MeV, 
while the total kinetic energy in the three-body cluster is typically more 
than three times larger and thus carried predominantly by the antikaon 
floating between the two slowly moving, heavy nucleons. The antikaon's 
potential energy roughly cancels the total kinetic energy, leaving room for 
the nucleon-nucleon interaction to bind the system which, as an isolated 
proton-proton pair, would be unbound. 

The admixture of the isospin-zero ($T_N=0$) component $|\Phi_-\rangle$ 
[Eq.~(\ref{Compont:-})] is typically about 4\%. It originates from the 
coupling matrix element 
$\langle \Phi_+ | \hat{V}_{\bar{K}N} | \Phi_- \rangle$ which is proportional
to the difference of $I=0$ and $I=1$ $\bar{K}N$ interactions. Although 
small, this admixture helps binding the $ppK^-$ system: without the 
$|\Phi_-\rangle$ component, the total binding energy would decrease by about
5 - 7 MeV as seen from the lines denoted ``HNJH$^\dagger$'' in 
Tables~\ref{tab:Res/Smooth/Wf1} and \ref{tab:Res/Smooth/Wf2}.

The spin-spin and $L^2$ parts of the $NN$ potential have an influence on the
$|\Phi_-\rangle$ admixture. The contribution of these terms to $V(NN)$ is
small and attractive. It tends to reduce the mixing of the 
$|\Phi_-\rangle$ component into the total wave function. This is seen in the
last lines of Tables~\ref{tab:Res/Smooth/Wf1} and \ref{tab:Res/Smooth/Wf2}
denoted by ``HNJH$^*$'' where the spin-spin and $L^2$ potentials are turned 
off and only the central part of the $NN$ potential is active. The 
spin-spin and $L^2$ terms were not considered in our previous 
Ref.~\cite{DHW:2008}.

\begin{table*}
\caption{\label{tab:Res/Smooth/Wf1} 
Detailed compilation of $ppK^-$ results calculated for chiral SU(3) models ORB, HNJH, BNW and BMN (see text) with the ``Type I'' ansatz for the
relation between antikaon binding energy $B_K$ and $\sqrt{s}$ in the off-shell $\bar{K}N$
two-body subsystem. The listing includes: the total binding energy $B(ppK^-)$ and decay width 
$\Gamma$ for  $\bar{K}NN\rightarrow\pi Y N$; the total kinetic energy, $E_{kin} =  \langle\Psi|\hat{T} - \hat{T}_{CM}|\Psi\rangle$, the nuclear part of the kinetic energy, $T_{nuc} =  \langle\Psi|\hat{T}_N - \hat{T}_{CM,N}|\Psi\rangle$; the antikaon binding energy $B_K$ as defined in Eq.(\ref{eq:BK}); the $\bar{K}$-nuclear potential energy, $V(\bar{K}N) = \langle\Psi|\text{Re}\,\hat{V}_{\bar{K}N}|\Psi\rangle$; the contribution from the nucleon-nucleon interaction, $V(NN) = \langle\Psi|\hat{V}_{NN}|\Psi\rangle$, with all entries given in MeV. Lower part of table: percentage $P(T_N=0)$ of the  $|\Phi_-\rangle$ admixture to the total wave function; r.m.s. distances $R_{NN}$ and $R_{\bar{K}N}$ between the 
two nucleons, and between antikaon and a nucleon; r.m.s.  
distances between antikaon and nucleon, $R_{\bar{K}N}$ $(I=0,\,1)$, for isospins $I = 0,1$ of the
$\bar{K}N$ pair (all given in fm). Lines denoted
HNJH$^\dagger$ are calculated switching off the admixture of the $|\Phi_-\rangle$ component. 
Lines HNJH$^*$ are results obtained without 
the spin-spin and $L^2$ potentials of the $NN$ potential. 
(Both these cases refer to the $\bar{K}N$ potential derived from the HNJH model). }

\begin{ruledtabular}
\begin{tabular}{lccccccc}
 & $B(ppK^-)$ & $\Gamma$ & $E_{kin}$ & $T_{nuc}$  & $B_K$ & $V(\bar{K}N)$ & $V(NN)$  \\
\hline 
ORB & 18.9 & 53.1 & 125.9 & 38.3 & 40.2 & $-$127.7 & $-$17.0 \\
HNJH & 16.9 & 47.0 & 129.5 &  38.1 & 38.9 & $-$130.4 & $-$16.2 \\
BNW & 18.1 & 60.4 & 124.9 &  37.4 & 39.2 & $-$126.7 & $-$16.3\\
BMN & 16.6 & 38.8 & 141.5 &  39.9 & 40.8 & $-$142.5 & $-$15.7 \\
\hline 
HNJH$^\dagger$ & 12.0 & 44.8 & 115.1 &  37.0 & 31.0 & $-$109.0 & $-$18.2 \\
HNJH$^*$ & 15.9 & 47.1 & 129.6 &  37.3 & 38.9 & $-$131.2 & $-$14.3\\
\end{tabular}

\vspace{0.3cm}

\begin{tabular}{lccccc}
 &  $P(T_N=0)$ &  $R_{NN}$ & $R_{\bar{K}N}$ & $R_{\bar{K}N}$ $(I=0)$ & $R_{\bar{K}N}$ $(I=1)$ \\
\hline
ORB &   3.4 \% &  2.15 & 1.93 &  1.79 & 2.28  \\
HNJH & 3.8 \% &  2.21 & 1.97 &  1.82 & 2.33  \\
BNW &  4.1 \% & 2.20 & 1.97 & 1.81 & 2.35  \\
BMN & 3.8 \% & 2.23 & 2.00 & 1.86 & 2.34   \\
\hline
HNJH$^\dagger$ & 0 \% & 2.13 & 2.01 & 2.01 & 2.01 \\
HNJH$^*$ & 4.5 \% & 2.26 & 2.00 & 1.83 & 2.39 \\
\end{tabular}
\end{ruledtabular}
\end{table*}


\begin{table*}
\caption{\label{tab:Res/Smooth/Wf2} 
Same as Table \ref{tab:Res/Smooth/Wf1},  but using the ``Type II'' ansatz for the
relation between antikaon binding energy $B_K$ and $\sqrt{s}$ in the off-shell $\bar{K}N$
two-body subsystem.}

\begin{ruledtabular}
\begin{tabular}{lccccccc}
 & $B(ppK^-)$ & $\Gamma$ & $E_{kin}$ & $T_{nuc}$  & $B_K$ & $V(\bar{K}N)$ & $V(NN)$  \\
\hline 
ORB & 22.7 & 64.2 & 136.0 & 40.9 & 46.1& $-$141.2 & $-$17.6 \\
HNJH & 20.8 & 58.3& 141.0 & 40.9 &45.1 & $-$145.2 & $-$16.6 \\
BNW & 18.1 & 71.4 & 132.0& 39.3 & 43.2 & $-$136.0 & $-$16.7\\
BMN & 21.7 & 53.1 & 158.4 &  43.8 & 49.3 & $-$163.9 & $-$16.2 \\
\hline 
HNJH$^\dagger$ & 13.8 & 51.8 & 121.3 &  38.7 & 33.9 & $-$116.5 & $-$18.5 \\
HNJH$^*$ & 19.8 & 58.6 & 141.5 &  40.3 & 45.2 & $-$146.4 & $-$14.9\\
\end{tabular}

\vspace{0.3cm}

\begin{tabular}{lccccc}
 &  $P(T_N=0)$ &  $R_{NN}$ & $R_{\bar{K}N}$ & $R_{\bar{K}N}$ $(I=0)$ & $R_{\bar{K}N}$ $(I=1)$ \\
\hline
ORB &   3.7 \% &  2.10 & 1.85 &  1.71 & 2.21  \\
HNJH & 4.1 \% &  2.15 & 1.89 &  1.73 & 2.26  \\
BNW &  4.1 \% & 2.16 & 1.91 & 1.75 & 2.29  \\
BMN & 4.1 \% & 2.15 & 1.89 & 1.73 & 2.25   \\
\hline
HNJH$^\dagger$ & 0 \% & 2.09 & 1.96 & 1.96 & 1.96 \\
HNJH$^*$ & 4.8 \% & 2.19 & 1.91 & 1.74 & 2.31 \\
\end{tabular}
\end{ruledtabular}
\end{table*}

\subsection{Structure of the $\bm{ppK^-}$ cluster}

Consider now the more detailed structure and characteristic sizes of the 
$ppK^-$ system as found in the present calculations. For this purpose we 
introduce density distributions of the $NN$ and $\bar{K}N$ pairs in the 
quasibound compound as functions of the respective nucleon-nucleon and 
antikaon-nucleon distances, as follows:
\begin{eqnarray}
\label{dens1}
\rho_{NN} (\bm{x}) & = & \langle \Psi | \; 
                                  \delta^3 \left({\bm{r}}_1 - {\bm{r}}_2 - \bm{x} \right) 
                              | \Psi \rangle~, 
                            \\
\rho_{\bar{K}N} (\bm{x}) & = & \langle \Psi | \; 
                               \frac{1}{2} \sum_{i=1,2} 
                                  \delta^3 \left({\bm{r}}_{\bar{K}} - {\bm{r}}_i - \bm{x} \right) 
                              | \Psi \rangle~,
\end{eqnarray}
where ${\bm{r}}_{1,2}$ refer to the two nucleons and ${\bm{r}}_{\bar{K}}$ to
the antikaon. Both densities are understood as being normalized to one. The 
projected density distributions for  $\bar{K}N$ pairs with specific isospin 
$I = 0,1$ are 
\begin{eqnarray}
&& \rho_{\bar{K}N}^I (\bm{x}) = \nonumber \\
&& \hspace{0.2cm} \langle \Psi | \; 
                               \frac{1}{2} \sum_{i=1,2} \hat{P}_I(\bar{K}N_i) \, 
                                  \delta^3 \left({\bm{r}}_{\bar{K}} - {\bm{r}}_i - \bm{x} \right) 
                              | \Psi \rangle~. 
                              \label{dens2}
\end{eqnarray}
with the isospin projectors $\hat{P}_I(\bar{K}N_i)$. 

With these distributions one can determine mean-square distances
\begin{eqnarray}
R_{NN}^2 & = & \int d^3 \bm{x} \; \bm{x}^2\rho_{NN} (\bm{x})~,\\
R_{\bar{K}N}^2 & = & \int d^3 \bm{x} \; \bm{x}^2\rho_{\bar{K}N} (\bm{x})~,\\
R^2_{\bar{K}N}(I) & = & N^{-1}_I\int d^3 \bm{x} \; \bm{x}^2\rho_{\bar{K}N}^I (\bm{x})~,
\label{dens3}
\end{eqnarray}
with $N_I = \int d^3 \bm{x} \; \rho_{\bar{K}N}^I (\bm{x})$. These average 
distances are summarized in the lower parts of 
Tables~\ref{tab:Res/Smooth/Wf1} and \ref{tab:Res/Smooth/Wf2}. The relatively
weak binding of the system implies rather large $NN$ distances, typically 
around $R_{NN} \simeq 2.2$ fm, while the average $\bar{K}N$ distances are 
slightly smaller, $R_{\bar{K}N} \simeq 1.9 - 2$ fm. When looked at 
separately in the $I = 0$ and $I=1$ channels of the $\bar{K}N$ subsystem, 
the significantly stronger attraction in the $I = 0$ component drags the 
antikaon closer to the nucleon than in the $I = 1$ component.

\begin{figure*}
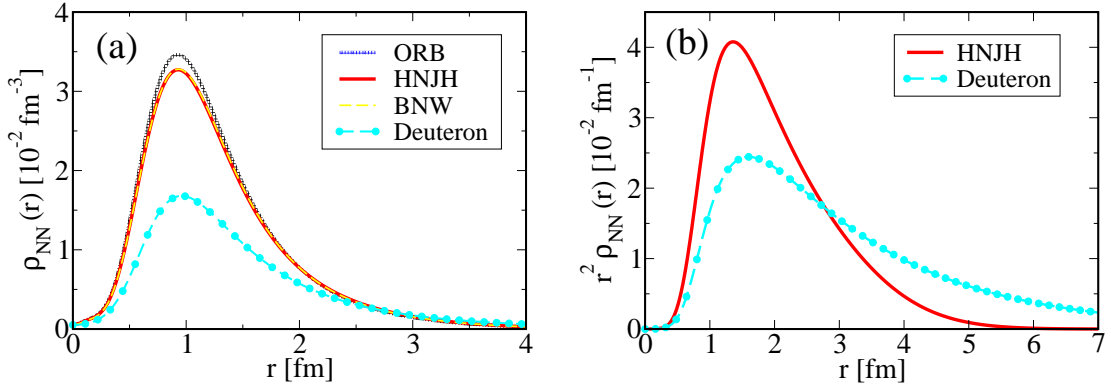

\includegraphics[width=7cm,angle=0,clip]{NNnoR2_v5.eps}
\hspace{0.5cm}
\includegraphics[width=7cm,angle=0,clip]{NNR2_v5.eps}

\caption{\label{fig:NNRelDens} (Color online) 
(a) $NN$ density in $ppK^-$, as function of $NN$ relative distance, calculated with the ``Type I'' ansatz.
``ORB'', ``HNJH'' and ``BNW'' results are shown as crossed, solid and dashed lines, respectively. 
For comparison, the deuteron density calculated with Av18 potential is depicted as dashed line with filled circle. 
(b) display of $r^2 \rho(r)$ for the same densities.}
\end{figure*}

\begin{figure*}
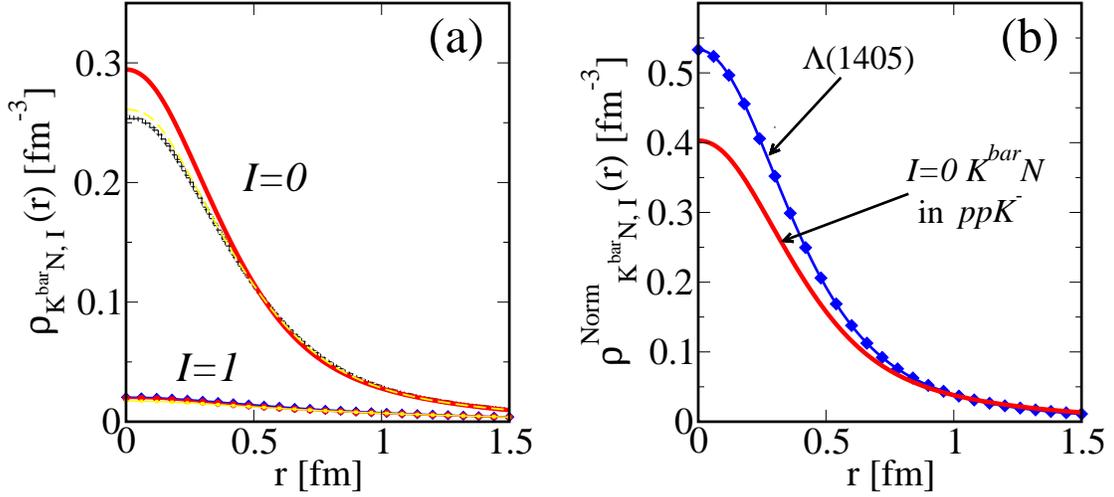

\includegraphics[width=7cm,angle=0,clip]{KNisoNoR2_v5.eps}
\hspace{0.5cm}
\includegraphics[width=7cm,angle=0,clip]{KNI0Norm_v5.eps}

\caption{\label{fig:KNRelDens} 
(Color online) 
$\bar{K}N$ density in $ppK^-$, as function of $\bar{K}N$ relative distance, calculated with the ``Type I'' ansatz. 
(a) 
Separate display of  each isospin component ($I=0$ and $I=1$). 
Lines without (with) symbols show the $I=0$ ($I=1$) component of the $\bar{K}N$ density.
Models ``ORB'', ``HNJH'' and ``BNW'' are depicted with crossed, solid and dashed lines, respectively. 
(b) 
Normalized $I=0$ $\bar{K}N$ relative density in $ppK^-$ 
for the model ``HNJH'' with the ``Type I'' ansatz 
(solid line). 
Solid line with diamond shows $\bar{K}N$ density in $\Lambda(1405)$ calculated with 
the same model. 
}

\end{figure*}

The different density distributions, Eqs.~(\ref{dens1}-\ref{dens2}), are 
shown in Figs.~\ref{fig:NNRelDens} and \ref{fig:KNRelDens}. These
densities depend only on the absolute value of the relative coordinate 
$\bm{x}$. The two-body densities are plotted as functions of $r = |\bm{x}|$ and 
normalized as $4\pi\int dr r^2 \rho(r)=1$. The two-nucleon distribution 
$\rho_{NN}$ shows the pronounced effect of the short-distance repulsive 
core of the $NN$ interaction. The maximum $NN$ density reached in this 
weakly bound system is about 0.03 fm$^{-3}$, a small fraction of typical 
bulk nuclear densities and only twice as large as the maximum proton-neutron
density in the even more dilute, very weakly bound deuteron.    

The $\bar{K}N$ density distribution has its maximum at zero distance between
the antikaon and each nucleon. It reflects the strong $\bar{K}N$ attraction 
in the $I=0$ channel, whereas the $I=1$ $\bar{K}N$ density is small. The 
right panel in Fig.~\ref{fig:KNRelDens} shows a comparison between densities
of the $I=0$ $\bar{K}N$ pair in the $ppK^-$ cluster and the one forming the 
$\Lambda(1405)$ as a two-body $\bar{K}N$ quasibound state. Both densities 
are properly normalized 
for the comparison. The conclusion to be drawn from 
this picture is that the $\Lambda(1405)$ stays essentially intact in the 
$ppK^-$ system which appears to behave much like a weakly bound, short-lived
$p\Lambda(1405)$ compound. Again, the $I=1$ component of the total 
$\bar{K}N$ density is small. A detailed analysis of the expectation values 
of the squared angular momentum, $L^2_{\bar{K}N}$ in the $ppK^-$ system 
shows that it is close to zero in $I=0$ and close to two in $I=1$ $\bar{K}N$
configurations, indicating as expected the dominant $s$-wave in the $I=0$ 
channel. The survival of the $\Lambda(1405)$ in the three-body 
cluster is qualitatively consistent with the result of the 
phenomenological potential~\cite{ppK:Akaishi}.

\subsection{Comparison with Faddeev results}

The small $ppK^-$ binding energy of only about 20 MeV found in the 
present variational approach appears to be inconsistent with results 
of a Faddeev calculation using a chiral interaction~\cite{Faddeev:Ikeda}
which predicts about 80 MeV binding. Here we discuss possible 
reasons for this difference.

An advantage of the Faddeev treatment is its capability to treat the 
three-body dynamics in the $\pi\Sigma N$ channel. The variational 
approach, on the other hand, works with an effective $\bar{K}N$ 
interaction after eliminating the $\pi\Sigma$ channel. While the 
attractive $\pi\Sigma$ two-body interactions and their 
coupled-channel effects are nonetheless accounted for as part of the 
complex and energy dependent $\bar{K}N$ potential, additional 
attraction may indeed be generated by the $\pi\Sigma N$ three-body 
dynamics treated explicitly in the Faddeev approach.

Secondly, there are significant differences in the subthreshold 
extrapolations of the two-body $\bar{K}N$ interaction. When solving 
Faddeev equations, a one-term separable form has been used to 
approximate the $\bar{K}N$ interaction~\cite{Faddeev:Ikeda}. While 
this interaction agrees with ours in its on-shell properties around 
$\bar{K}N$ threshold, it has been pointed out in 
Ref.~\cite{Hyodo:2007jq} that using such a separable approximation 
gives stronger attraction at lower (subthreshold) energies, as an artifact
of the regularization procedure. This is also evident by comparison 
of subthreshold extrapolations with early works on chiral SU(3) 
coupled-channel dynamics which had adopted separable forms for the 
$\bar{K}N$ and $\pi\Sigma$ interactions~\cite{Kaiser:1995eg}. In 
contrast, the $\bar{K}N$ potential in the present investigation is 
constructed using dimensional regularization of loops in the 
Bethe-Salpeter equation\footnote{This is equivalent to a dispersion 
relation approach, absorbing uncontrolled high-energy behavior in a 
few subtraction constants.} (see Appendix~\ref{App:ccSU(3)} for a further 
detailed assessment of different regularization schemes). The 
resulting subthreshold amplitudes have significantly smaller
real parts in comparison with those using the separable approximation.
The reason for this behavior can be traced to the three-momentum 
form factors commonly used in separable interactions. Standard 
analytic continuation of the $\bar{K}N$ momentum variable into the 
subthreshold region lets these form factors increase beyond their 
threshold (zero-momentum) magnitudes, thereby enhancing the 
subthreshold amplitude artificially. While such analytic 
continuations need not be performed in the Faddeev approach, this 
example nevertheless demonstrates that extrapolations into the 
far-subthreshold region are confronted with off-shell uncertainties 
which severely restrict the predictive power of $\bar{K}NN$ binding 
energy calculations.

Next, we examine the dispersive effect induced by the imaginary part of the 
$\bar{K}N$ potential. The advantage of the Faddeev method is its 
ability to deal consistently to all orders with the imaginary parts of the 
interactions whereas the variational method can handle this only 
perturbatively. In Section~\ref{subsec:dispersive} we have estimated that 
dispersive corrections from these imaginary parts amount to about $6\pm3$ 
MeV binding per nucleon, based on the analysis of the two-body $\bar{K}N$ 
channel. Thus the dispersive correction to the $ppK^-$ system would add 
another $\Delta B(ppK^-) \lesssim 15$ MeV to the total binding energy.

On top of these effects, there are several minor differences between two 
schemes such as the $NN$ potential, admixture of the $|\Phi_-\rangle$ 
component, and so on. These factors might work together to 
contribute to the difference between the present result and that in 
Ref.~\cite{Faddeev:Ikeda}.


We add a short comment on the sensitivity of the $ppK^-$ results to 
details of the $NN$ potentials. In fact, as long as the $\bar{K}NN$ 
system is only weakly bound, the dependence on different types of 
$NN$ interactions is weak. We have performed test calculations using a 
soft-core potential quite different from Av18 but equivalent with 
respect to reproducing low-energy $NN$ data. The resulting properties 
of the $ppK^-$ quasibound system turn out to be very similar to those 
obtained with the Av18 interaction. Details are given in
Appendix \ref{Sec:Minnesota}.


\subsection{Corrections from $\bm{p}$-wave $\bm{\bar{K}N}$ interactions}

\begin{figure}
\includegraphics[width=8cm,angle=0]{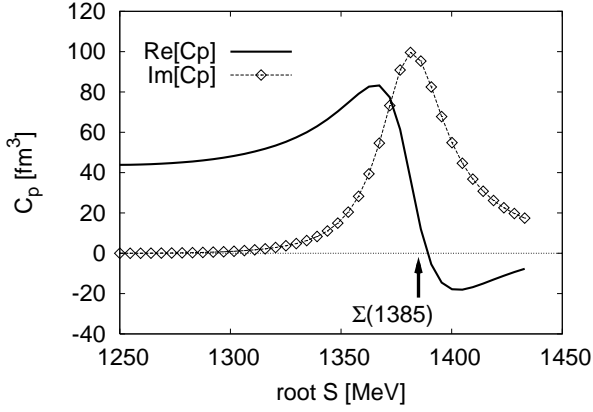}
\caption{\label{fig:ScattVol} 
Energy dependence of $K^-$-proton scattering volume, $C_p$. 
Its real (imaginary) part is drawn with a solid line (dotted line with diamond).}  
\end{figure}

A rough leading-order estimate of the effect of $p$-wave $\bar{K}N$ 
interactions can be performed by computing expectation values of suitably 
parametrized $p$-wave $K^-p$ and $K^-n$ potentials. 
The $p$-wave $\bar{K}N$ effective potentials used here, 
\begin{eqnarray}
v^{p\text{-wave}}_{\bar{K}N} (\bm{r}, \sqrt{s})  =  
 -  \frac{2 \,C_{N}(\sqrt{s})}{\sqrt{\pi}\, a_p^3\,\tilde{\omega}} \,\nabla \exp[-\bm{r}^2/a_p^2]\, \nabla~, \label{KNp}
\end{eqnarray}
with $\bm{r} = \bm{r}_{\bar K} - \bm{r}_N$ and the reduced $\bar{K}N$ energy
$\tilde{\omega}$, involve the energy-dependent $p$-wave scattering 
``volumes'' $C_p$ for $K^-$-proton and $C_n$ for $K^-$-neutron, where 
$C_n \simeq 2 \, C_p$. Their prominent feature is the $\Sigma(1385)$ 
resonance. We use a parametrization updated from Ref.~\cite{BWT78} which has
also been used in our previous study~\cite{DW-HYP06}. The energy dependence 
of $C_p$ is shown in Fig.~\ref{fig:ScattVol}. 

Expectation values $ \Delta V_{\bar{K}N} = 
\langle\Psi| v^{p\text{-wave}}_{\bar{K}N_1}
+ v^{p\text{-wave}}_{\bar{K}N_2}|\Psi\rangle$ of this $p$-wave $\bar{K}N$ 
potential are then computed with the $ppK^-$ wave functions of all four 
chiral models under consideration, changing the range parameter $a_p$ within
a reasonable interval, 0.4 fm to 0.9 fm. Then 
$\text{Re}\,\Delta V_{\bar{K}N}$ is found to be small and repulsive,
\begin{eqnarray}
1.5  \;  \lesssim \;  \text{Re}\,\Delta V_{\bar{K}N}\; \lesssim \; 5.0 \hspace{0.1cm} {\rm MeV}~.
\end{eqnarray}
The weak binding of the $ppK^-$ system places the effective $\sqrt{s}$ for 
the $\bar{K}N$ subsystem above the position of the $\Sigma(1385)$, hence the
positive sign of  $\text{Re}\,\Delta V_{\bar{K}N}$. The contribution of 
$p$-wave $\bar{K}N$ interactions is small 
if the $ppK^-$ binding from the driving $s$-wave interactions is relatively 
weak as it turns out in the present work. If the binding were strong enough 
to move the effective energy of the $\bar{K}N$ pair down below the 
$\Sigma(1385)$ resonance, the $p$-wave potential would act attractively and 
would tend to increase the binding. The present results appear to rule out 
this possibility.

For the correction to the $ppK^-$ width from $p$-wave $\bar{K}N$ 
interactions one finds 
$\Delta\Gamma = -2\,\text{Im}\,\Delta V_{\bar{K}N} \sim 10$ - 35 MeV. 
This relatively large $\Delta\Gamma$ (especially for the type I case) results from the fact that 
the two-body energy of the
$\bar{K}N$ subsystem is located close to the $\Sigma(1385)$ resonance with its prominent
imaginary part, see Fig. \ref{fig:ScattVol}. Obviously we can give here only a rough estimate for orientation.

\subsection{Estimate of antikaon absorption by the two-nucleon pair}

A further point of interest is the contribution to the decay width of the 
$ppK^-$ cluster from the two-body absorption process $K^{-}pp\to YN$. This 
effect is not included in the imaginary part of the $\bar{K}N$ potential 
used to estimate the decay width (\ref{width1}) in previous sections.

For first guidance, let us start with the formula of the decay width for 
$K^-$ absorption on proton pairs in a heavy nucleus~\cite{Mares:2006vk,
Friedman:2007zz,Weise:2008aj}
\begin{align}
   \Delta\Gamma_{abs} 
    = 
    \frac{2\pi  \tilde{B}_0}{\omega}
    \beta_{pp}(\omega)
    \int d^3\bm{r}\,
    \rho_{\bar{K}}(\bm{r})\,
    \rho_N^2(\bm{r})
    \label{eq:2Nabsorblarge} ,
\end{align}
where $\rho_N$ and $\rho_{\bar{K}}$ are the one-body densities of nucleon 
and antikaon, $\omega = m_K-B$ is the energy of the meson, and 
$\beta_{pp}(\omega)$ is a kinematical factor normalized to unity at 
threshold, $\omega = m_K$~\cite{Mares:2006vk,Weise:2008aj}. This factor 
reflects the phase space and kinematics for the relevant decay channels 
($\Sigma^+n$, $\Sigma^0p$, and $\Lambda p$ in the present case):
\begin{align}
    \beta_{pp}(\omega)
    =& \sum_{Y=\Lambda, \Sigma}
    \frac{\alpha_Y}{3}\left(\frac{M(m_K)}{M(\omega)}\right)^3
    \nonumber \\
    &\times \frac{\sqrt{[M^2(\omega)-m_N^2+M_Y^2]^2-4M^2(\omega)M_Y^2}}
    {\sqrt{[M^2(m_K)-m_N^2+M_Y^2]^2-4M^2(m_K)M_Y^2}}
    \nonumber
\end{align}
where $M(\omega)=2M_N+\omega$ and 
$\alpha_\Lambda=1$,  $\alpha_\Sigma =2$. 
Eq.~(\ref{eq:2Nabsorblarge}) derives from an effective contact interaction, 
with the coupling strength $\tilde{B}_0$ estimated empirically as 
$\tilde{B}_0\sim 1$ fm$^4$ \cite{Weise:2008aj} on the basis of kaonic atom data
summarized and discussed in Refs. \cite{Mares:2006vk,Friedman:2007zz}.

The absorptive width in the form (\ref{eq:2Nabsorblarge}) is proportional to
the probability of finding three particles (antikaon and two nucleons) at 
the same space point, as expressed by the product of the one-body densities.
This treatment is justified for large nuclei where the independent particle 
picture works reasonably well. However, for the few-body $K^-pp$ system, the
following modifications are required:
\begin{itemize}
    \item[(i)] correlations between the two nucleons must be taken into 
    account, and 

    \item[(ii)]  finite range effects in the absorption process must be 
    considered.
\end{itemize}
The first modification is mandatory because of the repulsive core in the 
nucleon-nucleon interaction. Once the correlations are taken into account, a
local absorptive contact interaction~\eqref{eq:2Nabsorblarge} is not 
appropriate because it provides almost no width: the probability for two 
nucleons to be found at the same spot is basically zero. A proper treatment 
must therefore deal with finite range effects in the absorption process 
which in turn requires a more detailed assessment of the underlying 
microscopic mechanisms. 

\begin{figure}[tbp]
    \centering
    \includegraphics[width=0.5\textwidth,clip]{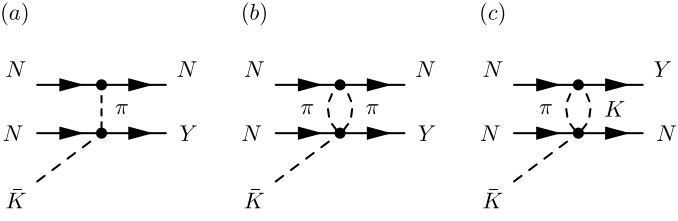}
    \caption{\label{fig:micro}
    Feynman diagrams for typical microscopic mechanisms of the two-nucleon 
    absorption process. (a): one pion exchange which is forbidden in the 
    present system, (b) and (c): two-meson 
    exchange. 
    }
\end{figure}

By analogy with pion absorption in the deuteron, a leading microscopic 
process would be one-pion exchange (Fig.~\ref{fig:micro}, left). This 
process, however, does not contribute to $K^-$ absorption on a proton-proton
pair with spin $S=0$ in the $K^-pp$ system considered here. So the driving 
absorption mechanism is expected to come from exchanges of two pseudoscalar 
mesons as illustrated in Fig.~\ref{fig:micro}. In chiral effective field 
theory, these are subleading one-loop terms in the two-baryon system 
involving $\bar{K}\pi\pi$ or $\bar{K}K\pi$ couplings to one of the baryons 
and exchange of the (interacting) two mesons with the second baryon. These 
processes occur when the external antikaon overlaps with one of the 
nucleons. The effective range of the absorption process is then related to 
the mass spectrum representing the exchanged two-meson system. 

Thus, in order to estimate the width from the two-nucleon absorption 
process, we need to extend the formula~\eqref{eq:2Nabsorblarge} to satisfy 
the requirements (i) and (ii). The result should reduce to the original 
form~\eqref{eq:2Nabsorblarge} in the limit of no $NN$ correlations and 
zero-range interaction, so that contacts to phenomenology can be made 
concerning the absorptive coupling strength $\tilde{B}_0$. 

First note that the product of one-body densities can be expressed as 
\begin{align}
    \rho_{\bar{K}}(\bm{r})\, 
    \rho_N^2(\bm{r}) 
    = &
    \int d^3\bm{x}_1 \int d^3\bm{x}_2\, 
    \rho_{\bar{K}}(\bm{r})\,
    \rho_N(\bm{x}_1)\,\rho_N(\bm{x}_2)  \nonumber \\ 
    &\times\delta^3(\bm{x}_1-\bm{x}_2)
    \sum_{i=1,2}\frac{\delta^3(\bm{x}_i-\bm{r})}{2} ~. 
    \label{eq:density}
\end{align}
For the modification requested by item (i), we introduce the three-body 
density as 
\begin{align}
   &\rho^{(3)}(\bm{r},\bm{x}_1,\bm{x}_2) 
   \nonumber \\
   \equiv &\langle \, \Psi \, | 
    \delta^3(\bm{r}_{\bar{K}}-\bm{r}) 
    \delta^3(\bm{r}_1-\bm{x}_1) 
   \delta^3(\bm{r}_2-\bm{x}_2) 
    | \, \Psi \, \rangle \nonumber \\ 
    = &
    \rho_{\bar{K}}(\bm{r}) 
    \rho_N(\bm{x}_1)\rho_N(\bm{x}_2) 
    [1-C(\bm{r},\bm{x}_1,\bm{x}_2)]~,
    \nonumber
 \end{align}
where the function $C(\bm{r},\bm{r}_1,\bm{r}_2)$ represents the correlations
among the particles. The product of one-body densities in 
Eq.~\eqref{eq:density} can now be replaced by the three-body density 
$\rho^{(3)}(\bm{r},\bm{x}_1,\bm{x}_2) $ which reduces to the original form 
in the limit of $C\to 0$. The modification (ii) is implemented by replacing 
the delta function in Eq.~\eqref{eq:density} by a finite range distribution,
such as a normalized Gaussian: 
\begin{align}
    \delta^3(\bm{x}_1-\bm{x}_2) 
    \to G(\bm{x}_1-\bm{x}_2;a) 
    =\frac{1}{\pi^{3/2}a^3}e^{-|\bm{x}_1-\bm{x}_2|^2/a^2}~. 
    \nonumber
\end{align}
Taking $a\to 0$, this distribution turns into the delta function. In 
summary, the formula for the absorptive width of the $ppK^-$ few-body system
is given by 
\begin{align}
    \Delta\Gamma_{abs}& (K^{-}pp\to YN) 
    = 
    \frac{2\pi  B_0}{\omega}
    \beta_{pp}(\omega)\nonumber \\
    &\times \int d^3\bm{r}\int 
    d^3\bm{x}\,\,
    \rho^{(3)}(\bm{r},\bm{r},\bm{x})\, 
    G(\bm{x}-\bm{r};a)~.
    \label{eq:2Nabsorb} 
\end{align}

\begin{figure}[t]
\includegraphics[width=8cm,angle=0]{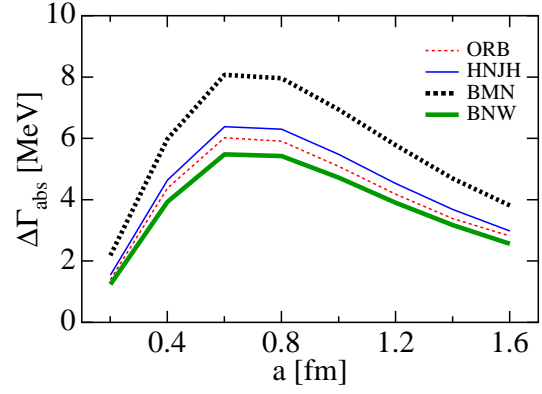}
\caption{(Color online) Absorptive width $\Delta\Gamma_{abs}$ for $ppK^- \rightarrow YN$
as function of Gaussian range parameter $a$. Results are shown for
the models ``ORB'', ``HNJH'', ``BNW'' and ``BMN'' discussed in the text.} 
\label{fig:gamma2N} 
\end{figure}

The variation of $\Delta\Gamma_{abs}$ with respect to changes of the 
Gaussian range parameter $a$ is shown in Fig.~\ref{fig:gamma2N}. Note that, 
from the point of view of the microscopic $\pi\pi$ or $\pi K$ exchange 
mechanisms discussed previously, the characteristic ranges are covering a 
band from $a\sim 0.4$ fm for $K^*$ exchange to about $a\sim 1.4$ fm for 
uncorrelated $2\pi$ exchange. For a typical choice of the absorptive 
coupling strength, $\tilde{B}_0 = 1$ fm$^4$ and $a = 0.6$ fm, one finds: 
\begin{equation} 
5~ {\rm MeV}\, \lesssim \, \Delta\Gamma_{abs} \, \lesssim\, 8~ {\rm MeV} 
\end{equation} 
for all models (with Type I or Type II ansatz). The coupling strength $\tilde{B}_0$ is, of course, subject
to considerable uncertainties. From the recent 
survey~\cite{Friedman:2007zz} based on kaonic atom data 
analysis one estimates a possible band  width $0.85 < \tilde{B}_0 <1.5$ fm$^4$ \cite{Weise:2008aj}
which translates into
\begin{equation} 
4 \; {\rm MeV} \lesssim \; \Delta\Gamma_{abs}  \; \lesssim\; 12 \; {\rm MeV}~.
\end{equation}
Given this estimated upper limit for $\Delta\Gamma_{abs}$ in the $ppK^-$ system we 
conclude that the absorptive width in such a dilute, weakly bound system is 
expected to be less important than the one from the mesonic 
$\bar{K}NN\rightarrow \pi Y N$ decay under the given conditions. This 
situation may change qualitatively for heavier nuclei with potentially 
stronger $\bar{K}$ binding where the absorptive width can be far more 
prominent as pointed out in Ref.~\cite{Weise:2008aj}.

\section{SUMMARY}

We have investigated the prototype of antikaonic nuclei, $ppK^-$, using a 
variational approach with an effective $\bar{K}N$ interaction derived from 
chiral SU(3) coupled-channel dynamics. Several versions of such $\bar{K}N$ 
interactions have been employed in the calculations. They all satisfy the 
necessary constraint of reproducing empirical $K^-p$ threshold information 
and the $\pi\Sigma$ mass spectrum in the region of the $\Lambda(1405)$, 
within (admittedly large) uncertainties of the existing experimental data 
base. Furthermore, a realistic nucleon-nucleon interaction (Argonne v18) has
been used throughout. This interaction properly accounts for repulsive 
short-distance $NN$ correlations.

The primary problem faced in the theoretical part of the quest for 
antikaon-nuclear quasibound systems is the subthreshold extrapolation of the
$s$-wave $\bar{K}N$ interaction. As a consequence of the strong 
$\bar{K}N\leftrightarrow \pi\Sigma$ coupled-channel dynamics, this 
interaction is complex, non-local and energy dependent. While the driving 
interaction kernel is determined by chiral SU(3) dynamics, the off-shell 
extrapolation into the far-subthreshold region is subject to uncertainties 
which limit the predictive power of the theory. A minimal condition for any 
such calculation is to account in detail for the coupled-channel dynamics 
that governs the formation of the $\Lambda(1405)$ as a quasibound $\bar{K}N$
state embedded in the strongly interacting, resonant $\pi\Sigma$ continuum. 
When this is done, the resulting $\bar{K}N$ effective potential turns out to
be significantly less attractive than anticipated in a simple 
phenomenological approach using a local, energy-independent potential. As a
consequence we arrive at weaker $ppK^-$ binding than that previously 
suggested.

As an independent test of the variational method applied in the present 
investigation, we have performed calculations using the phenomenological 
Akaishi-Yamazaki (AY) potential. The stronger binding found with this energy
independent local potential is indeed reproduced as reported in detail in 
Appendix~\ref{Sec:Bench}.

The results of the present variational calculation 
are summarized as follows:
\begin{itemize}

\item{The calculated binding energy of the $ppK^-$ cluster,
based on the leading $s$-wave 
$\bar{K}N$ interaction only, is
\begin{equation}
B(ppK^-) \simeq 20 \pm 3~\text{MeV}~,
\nonumber
\end{equation}
where the error indicates variations using four different versions of chiral
SU(3) coupled-channel calculations. The decay width into $\pi Y N$ final 
states is estimated to be in the range
\begin{equation}
\Gamma(ppK^-\rightarrow\pi Y N) \sim  40~\text{-}~70~\text{MeV}~.
\nonumber
\end{equation}}


\item{Differences between these variational results and those obtained from 
Faddeev calculations~\cite{Faddeev:Ikeda} presumably relate in large part to
$\pi\Sigma N$ three-body dynamics not incorporated in the present framework.
Some part of these differences may be attributed to the use of separable
approximations for the $\bar{K}N$ and $\pi Y$ interactions when solving 
Faddeev equations, and to dispersive effects not covered by the 
variational approach. We estimate the dispersive correction to the total 
$ppK^-$ binding energy to be $\Delta B(ppK^-) \lesssim 15$ MeV.}

\item{This additional binding is partly compensated by corrections from 
$p$-wave $\bar{K}N$ interactions which have a repulsive effect as long as 
the effective two-body energy in the $\bar{K}N$ subsystem stays above the 
$\Sigma(1385)$ resonance. The overall binding energy, after corrections and 
with conservative error assignment, is then expected to be in the interval 
$B(ppK^-) \simeq 20$~-~40 MeV. }  

\item{The $p$-wave interactions tend to increase the mesonic decay width for
$ppK^- \rightarrow \pi Y N$ by an amount of 10 - 35 MeV. } 

\item{The additional effects of $\bar{K}NN\rightarrow YN$ absorption are 
estimated to increase the decay width of the quasibound $ppK^- $ state 
further by an mount of order 10 MeV}. 

\item{In the variational approach the wave function of the $ppK^-$ 
quasibound state can be computed and analyzed. Given its weak 
binding, the system is rather dilute. The average distance between the two 
nucleons is about $2.1 - 2.2$ fm. The short-range repulsion in the $NN$ 
interaction plays an important role in keeping the nucleons apart. The 
antikaon likes to minimize its distance from either nucleon. The isospin 
$I = 0$ $\bar{K}N$ density distribution within the $\bar{K}NN$ cluster is 
reminiscent of the $\Lambda(1405)$ in the $\bar{K}N$ two-body system, 
indicating a $\Lambda(1405)N$ dibaryonic hybrid structure. However, with its
estimated lifetime of order $\tau = 1/\Gamma \sim 2$ fm/c, this structure 
can exist only over a very short time interval.}

\end{itemize}

Taking all theoretical uncertainties into account, we arrive at the 
conclusion that narrow, deeply bound $ppK^-$ clusters are unlikely to exist.
Our calculation predicts such systems to be weakly bound and short-lived. 
With a rather low binding energy and a total width between 60 and 120 
MeV, such structures would indeed be difficult to detect. One should keep in
mind, however, that the present calculations rely entirely on the 
constraints provided by the presently available sets of $K^-p$ threshold 
data and the poorly known $\pi\Sigma$ mass spectrum. Improvements of this 
data base will sharpen the theoretical input conditions for deriving the 
effective subthreshold $\bar{K}N$ interaction. One can look forward to 
further developments along these lines as the experimental searches proceed 
with higher precision.

\section*{ACKNOWLEDGEMENTS}

We thank Avraham Gal for fruitful and stimulating discussions.
One of authors (A. D.) is grateful to Prof. Akaishi 
for advice on the construction of our model wave function.
This project is partially supported by BMBF, GSI and by the DFG excellence cluster 
``Origin and 
Structure of the Universe".  T.~H. thanks the Japan Society for the Promotion
of Science (JSPS) for financial support. This work is also supported in part by 
the Grant for Scientific Research (No.\ 19853500, 19740163) from the Ministry
of Education, Culture, Sports, Science and Technology (MEXT) of Japan. This 
research is  part of Yukawa International Program for Quark-Hadron Sciences.

\appendix

\section{NN potentials}\label{App:NN}

\begin{figure}
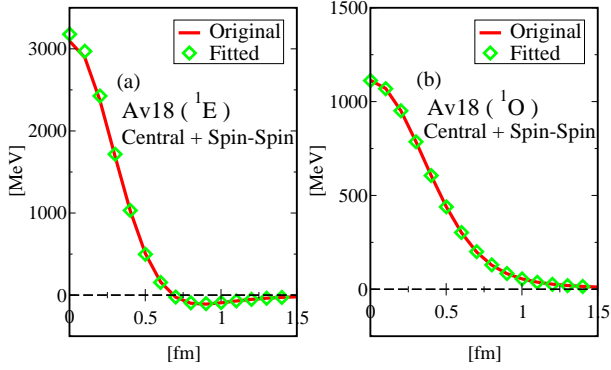

\includegraphics[width=4cm,angle=0,clip]{Av18_1E_CSS_v3.eps}
\includegraphics[width=4cm,angle=0,clip]{Av18_1O_CSS_v3.eps}
\caption{\label{fig:Av18_CSS} (Color online) Central plus spin-spin $NN$ potentials in singlet-even (a)
and singlet-odd (b) channels. Solid line: original Argonne v18 potential (``Original''). 
Points represent fits to the original potential with 
four Gaussians (``Fitted'').}
\end{figure}

\begin{figure}
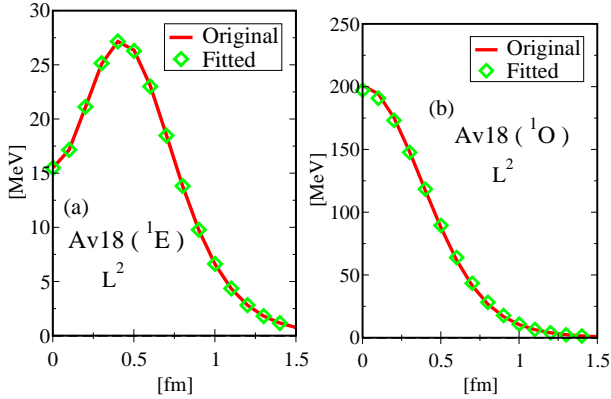

\includegraphics[width=4cm,angle=0,clip]{Av18_1E_L2_v3.eps}
\includegraphics[width=4cm,angle=0,clip]{Av18_1O_L2_v3.eps}
\caption{\label{fig:Av18_L2} (Color online) $L^2$ potential in $NN$ potential. 
Figs. (a) and (b) show $^1E$ and $^1O$ channels, respectively. 
Solid line is the original Argonne v18 potential (``Original''). 
Points represent the potential used in the present study, which fit 
the original potential with three Gaussians (``Fitted'').}
\end{figure}

\begin{table}[t]
\caption{\label{tab:Av18fitted} 
Best fit parameters for the expansion of the Av18 potential into Gaussians. Strengths
are given in MeV, ranges in fm.}
\begin{ruledtabular}
\begin{tabular}{lrrrr}
$n$                          & 1 & 2 & 3 & 4  \\
\hline \\
$v^{c+SS, 1E}_n$ & 3708  & $-$483 & $-$44  & $-$3   \\
$r^{c+SS, 1E}_n$ & 0.411 & 0.726  & 1.317  & 2.619  \\
\\
$v^{c+SS, 1O}_n$ & 651   & 412    & 43     & 6  \\
$r^{c+SS, 1O}_n$ & 0.453 & 0.603  & 1.092  & 2.622  \\
\\
$v^{L2, 1E}_n$   & $-$47 & 57     & 5      &   \\
$r^{L2, 1E}_n$   & 0.383 & 0.637  & 1.009  &   \\
\\
$v^{L2, 1O}_X$   &    170 & 27  & $-$0.3  &   \\
$r^{L2, 1O}_X$   &    0.537 & 0.804  & 1.644  &  \\
\\    
\end{tabular}
\end{ruledtabular}
\end{table}

For practical convenience in performing the variational calculations, the original Av18 potential has
been represented by a series of Gaussians. 

The spin state of the two nucleons is restricted to $S=0$ in our model, 
so the expectation value of the operator $\hat{\bm{\sigma}}_1 \cdot \hat{\bm{\sigma}}_2$ 
in the spin-spin potential is equal to $-3$. 
We have fitted the radial parts of the central potential plus the spin-spin potential, 
$v^c_{X} (r) -3 v^{SS}_{X} (r)$, with four Gaussians of different ranges; 
the radial part of $L^2$ potential, $v^{L2}_{X} (r)$, is fitted with three Gaussians:   
\begin{eqnarray}
v^c_{X} (r) -3 v^{SS}_{X} (r) & = &  \sum_{n=1}^4 v^{c+SS, X}_n \; e^{- (r / r^{c+SS, X}_n )^2} \, , \\
v^{L2}_{X} (r) & = & \sum_{n=1}^3 v^{L2, X}_n \; e^{-(r / r^{L2, X}_n )^2} \, ,    
\end{eqnarray}
with strengths $v^{c+SS, X}_n$ and $v^{L2, X}_n$ and ranges
$r^{c+SS, X}_n$ and $r^{L2, X}_n$, where $X$ indicates the channels 
$^1E$, $^1O$.  
The values of the parameters are listed in Table \ref{tab:Av18fitted}. 
For illustration of the fit quality we draw the potentials in 
Figs.~\ref{fig:Av18_CSS} and \ref{fig:Av18_L2}. 

\section{Chiral coupled-channel approach and regularization 
dependence}\label{App:ccSU(3)}

Here we briefly summarize basics of the chiral SU(3) coupled-channel 
approach~\cite{Hyodo:2007jq}. In particular, we study the regularization 
dependence of the $\bar{K}N$ and $\pi\Sigma$ amplitudes, an issue that has 
repeatedly been raised in the literature. 

The starting point is the coupled-channel meson-baryon scattering equation 
\begin{equation}
    T_{ij} = V_{ij} + V_{ik}\,G_k\,T_{kj} ,
    \label{eq:BS}
\end{equation}
with the interaction kernel $V_{ij}$ derived from the chiral SU(3)
meson-baryon Lagrangian, and the loop function $G_k$ to be discussed in 
detail later. In order to derive the single-channel $\bar{K}N$ potential, 
an effective interaction kernel $V^{\text{eff}}$ is constructed such that the full 
solution of Eq.~\eqref{eq:BS} in the $\bar{K}N$ channel is reproduced by 
solving a single-channel scattering equation with $V^{\text{eff}}$:
\begin{equation}
    T_{11} = V^{\text{eff}} + V^{\text{eff}}\,G_{1}\,T_{11} ,
    \label{eq:BSsingle}
\end{equation}
where $i=j =1$ now represents the $\bar{K}N$ channel. This $V^{\text{eff}}$ is 
used in Eq.~\eqref{eq:uncorrected} to determine the strength of the 
$\bar{K}N$ potential.

In Ref.~\cite{Hyodo:2007jq}, dimensional regularization was used in the 
calculation of the loop function $G$. Dimensional regularization has the 
advantage that the analyticity of the loop function is compatible with the 
dispersion relation used in the N/D method~\cite{Oller:2000fj}. It is 
however instructive to study the regularization dependence in this 
framework, in order to understand possible differences between present 
results for the $K^-pp$ system and the results of the Faddeev calculation 
with interactions constrained by chiral dynamics~\cite{Faddeev:Ikeda}. An 
analysis of the regularization dependence in this framework can be found in 
Ref.~\cite{Nam:2003ch}. Note that differences in the regularization schemes 
only change the loop function $G$ and leave the interaction kernel $V$ 
untouched.

Using dimensional regularization, the loop function $G$ is given by
\begin{align}
    G^{dim}(\sqrt{s})
    =&\frac{2M}{(4\pi)^{2}}
    \Biggl\{a(\mu)+\ln\frac{mM}{\mu^{2}}
    +\frac{\Delta}{s}\ln\frac{M}{m} \nonumber \\
    &+\frac{\bar{q}}{\sqrt{s}}
    \ln{
    \frac{\phi_{++}(s)\,\phi_{+-}(s)}{\phi_{-+}(s)\,\phi_{--}(s)}
    }
    \Biggr\} ,
    \label{eq:dimreg}
\end{align}
where $a(\mu)$ are subtraction constants, $\mu$ is the renormalization 
scale, and we have defined
\begin{align}
    \Delta 
    &=  M^{2}-m^{2} , \nonumber \\
    \phi_{\pm\pm}(s) 
    &= \pm s\pm \Delta+2\bar{q}\sqrt{s}~~,
    \nonumber\\
    \bar{q} 
    &= \frac{\sqrt{(s-(M-m)^2)(s-(M+m)^2)}}{2\sqrt{s}}~.
    \nonumber
\end{align}
With a sharp three-momentum cutoff, the loop function is
\begin{align}
    G^{3d}(\sqrt{s})
    =& \frac{2M}{(4\pi)^2}
    \Biggl\{
    \ln \frac{mM}{\qmax^2}
    +\frac{\Delta}{s}
    \ln \frac{M(1+\xi^{m})}
    {m(1+\xi^{M})} \nonumber \\
    &-\ln
    \left[
    \left(1+\xi^{m}\right)
    \left(1+\xi^{M}\right)
    \right]
    \nonumber \\
    &
    +\frac{\bar{q}}{\sqrt{s}}
    \ln \frac{\phi^m_+(s)\phi^M_+(s)}
    {\phi^m_-(s)\phi^M_-(s)}
    \Biggl\}~,
    \label{eq:3dcutana}
\end{align}
where $q_{max}$ is the three-momentum cutoff and 
\begin{align}
    \phi^{m}_{\pm}(s) 
    =& \pm s\mp \Delta
    +2\bar{q}\sqrt{s}\,\xi^{m}~, \nonumber \\ 
    \phi^{M}_{\pm}(s) 
    =& \pm s\pm \Delta
    +2\bar{q}\sqrt{s}\,\xi^{M},
    \nonumber \\
    \xi^{m}
    =& \sqrt{1+\frac{m^2}{\qmax^2}}~,
    \quad 
    \xi^{M}
    = \sqrt{1+\frac{M^2}{\qmax^2}}~.
    \nonumber 
\end{align}
A smooth cutoff can be introduced using the Pauli-Villars method which 
corresponds to multiplying a monopole form factor
\begin{equation}
    \frac{m^2-\Lambda^2}{q^2-\Lambda^2}
    \nonumber
\end{equation}
to the loop function, leading to
\begin{align}
    G^{PV}(\sqrt{s})
    =&\frac{2M}{(4\pi)^{2}}
    \Biggl\{\ln\frac{m}{\Lambda}
    +\frac{\Delta}{s}\ln\frac{M}{m}
    -\frac{\Delta_{\Lambda}}{s}\ln\frac{M}{\Lambda}
    \nonumber \\
    &+\frac{\bar{q}}{\sqrt{s}}
    \ln{
    \frac{\phi_{++}(s)\,\phi_{+-}(s)}{\phi_{-+}(s)\,\phi_{--}(s)}
    } \nonumber \\
    &-\frac{\bar{q}_{\Lambda}}{\sqrt{s}}
    \ln{
    \frac{\phi_{\Lambda,++}(s)\,\phi_{\Lambda,+-}(s)}
    {\phi_{\Lambda,-+}(s)\,\phi_{\Lambda,--}(s)}
    }
    \Biggr\}~ ,
    \label{eq:PV} 
\end{align}
where 
\begin{align}
    \Delta_{\Lambda}
    &=  M^{2}-\Lambda^{2} , \nonumber \\
    \phi_{\Lambda,\pm\pm}(s) 
    &= \pm s\pm \Delta_\Lambda+2\bar{q}_\Lambda\sqrt{s}~~.
    \nonumber \\
    \bar{q}_{\Lambda}
    &= \frac{\sqrt{(s-(M-\Lambda)^2)(s-(M+\Lambda)^2)}}{2\sqrt{s}} 
    \nonumber
\end{align}

\begin{figure}[tbp]
    \centering
    \includegraphics[width=0.5\textwidth,clip]{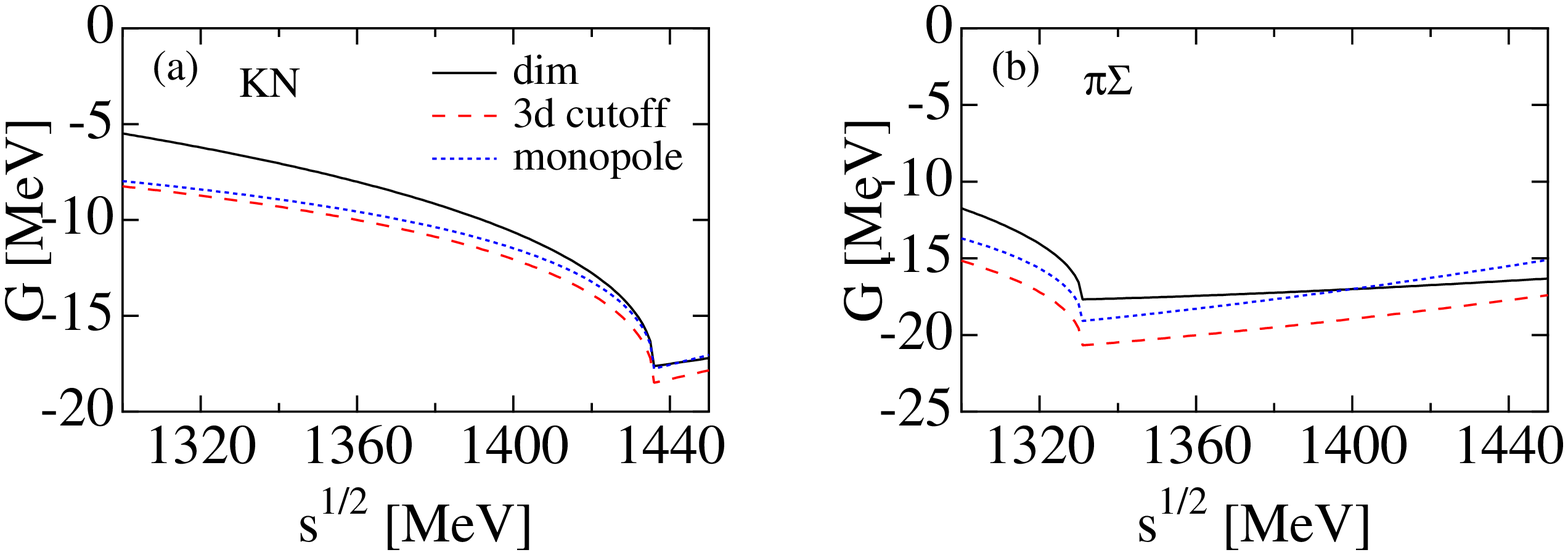}
    \caption{\label{fig:loopcomp}
    (Color online) Comparison of the loop functions: 
    dimensional regularization (solid lines),
    thee-momentum sharp cutoff (dashed lines),
    and
    smooth monopole form factor (dotted lines).
    (a): $\bar{K}N$ channel, (b):, $\pi\Sigma$ channel.}
\end{figure}

The real parts of the loop functions~\eqref{eq:dimreg}, \eqref{eq:3dcutana},
and \eqref{eq:PV} for $\bar{K}N$ and $\pi\Sigma$ channels are plotted in 
Fig.~\ref{fig:loopcomp}. The imaginary parts are determined by the phase 
space of intermediate meson-baryon states and therefore independent of the
regularization procedure. The parameters for the dimensional regularization 
and three momentum cutoff schemes are taken from phenomenologically 
successful models~\cite{Oset:1998it,Oset:2001cn}:
\begin{align*}
    a_{\bar{K}N} =& -2~,\quad a_{\pi\Sigma} = -1.84~,
    \quad \mu = 630 \text{ MeV} ~,
    \nonumber \\
    \qmax =& \, 630 \text{ MeV} .
    \nonumber 
\end{align*}
The loop functions $G^{dim}$ and $G^{3d}$ are evidently quite similar. 
Furthermore, choosing the parameter of the monopole form factor as
\begin{align*}
    \Lambda_{\bar{K}N} =& 750\text{ MeV} ,\quad 
    \Lambda_{\pi\Sigma} = 500 \text{ MeV} 
    \nonumber 
\end{align*}
the corresponding loop functions $G^{PV}$ behave quantitatively similar as 
those with dimensional/three-momentum cutoff schemes, as seen in 
Fig.~\ref{fig:loopcomp}. The cutoff scales of the form factors (several 
hundreds of MeV) is typical and naturally expected from meson-baryon 
phenomenology.

In summary, we find that smooth cutoff schemes provide similar loop 
functions as those with sharp cutoff or dimensional regularization. Since 
differences in the regularization schemes only affect the loop functions $G$
in the present framework, it is clear that the potentials derived in 
Ref.~\cite{Hyodo:2007jq} remain unchanged when adopting a smooth cutoff.

In this respect, the difference between the present results for $K^-pp$ 
binding and those in Faddeev method~\cite{Faddeev:Ikeda} cannot be 
attributed to the regularization method. It is worth noting that early 
studies using the coupled-channel approach~\cite{Kaiser:1995eg} gave a 
stronger $\bar{K}$ subthreshold amplitude than the one derived in 
Ref.~\cite{Hyodo:2007jq}. The calculations of Ref.~\cite{Kaiser:1995eg} used
a separable approximation for the potential together with the 
non-relativistic Lippmann-Schwinger equation, a feature also shared by the 
Faddeev calculations of Ref.~\cite{Faddeev:Ikeda}. Thus the larger binding 
energy of $K^-pp$ found in Ref.~\cite{Faddeev:Ikeda} is likely to be related
to the type of scattering equation and the separable approximation used in 
that work.

\section{Benchmark test with the AY potential \label{Sec:Bench}}

As a test for the variational method and trial wave functions applied in the present work,
we perform a calculation with the $\bar{K}N$ interaction [the Akaishi-Yamazaki 
(AY) potential] and the $NN$ interaction (the Tamagaki potential) used in Ref.\cite{ppK:Akaishi}. 
The AY potential is an energy-independent, local potential based on phenomenology but
not constrained by chiral SU(3) dynamics.

The results are summarized in Table \ref{tab:Bench}. 
The first two columns (``Model I'' and ``Model II'') 
show our results; the last column (``ATMS'') is the 
original result reported in Ref.~\cite{ppK:Akaishi}. 
In ``Model I'' the energy variation is performed restricting the wave function to its dominant 
$|\Phi_+\rangle$ component, i.e. with no admixture of the $|\Phi_-\rangle$ state (the one with
$NN$ coupled to isospin $T_N = 0$), keeping the coefficient $C\equiv 0$ in Eq.(\ref{Mix}). 
``Model II'' includes the  $|\Phi_-\rangle$ component, with $C$ determined variationally.
In all calculations, the convergence of the Gaussian expansions (\ref{NNcorr}) and (\ref{KNcorr})
has been checked and found satisfactory with $N_N=N_K=9$. 

\begin{table}
\caption{\label{tab:Bench} 
Test of the present variational calculation with the AY potential \cite{ppK:Akaishi}.
``Model I'' and ``Model II'' are the results using the variational wave function described
in Sec.~\ref{SS:Model}. ``ATMS'' refers to the results quoted in Ref.~\cite{ppK:Akaishi}. 
$B(ppK^-)$ and $\Gamma$ are the total $ppK^-$ binding energy and the  
$\bar{K}NN\rightarrow\pi Y N$ decay width. The antikaon binding energy $B_K$ is defined in Eq.~(\ref{eq:BK}). All the remaining
quantities are specified as in Table~\ref{tab:Res/Smooth/Wf1}.
All energies are given in units of MeV.  The r.m.s. distances $R_{NN}$ and  $R_{\bar{K}N}$ are 
in fm. }
\begin{ruledtabular}
\begin{tabular}{lrrrr}
          & Model I & Model II & ATMS \\
$C$ & 0       & finite   & ---  \\
\hline\\
$B(ppK^-)$ & 39.0  &   51.4   &  48 \\
$\Gamma$  & 60.0  &   61.0   &  61 \\
\\
$B_K$      & 65.8  &   80.0   &  68 \\
$T_{nuc}$ & 46.7  &   47.8   &  \\
\\
$E_{kin}$        & 147.0 &   162.4  & 167 \\
$V(NN)$       & $-$19.8  & $-$19.2  & $-$19 \\
$V(\bar{K}N)$     & $-$166.2 & $-$194.6 & $-$196 \\
\\
$R_{NN}$    & 1.75  & 1.83     & 1.90 \\
$R_{\bar{K}N}$     & 1.54  & 1.55     & 1.57 \\
\\
$P(T_N=0)$  & 0 & 5.9 \%   & --- \\
\\
\end{tabular}
\end{ruledtabular}
\end{table}

One evidently finds a high degree of consistency between ``Model II'' and ``ATMS'',
confirming that the different variational methods used here and in 
Ref.~\cite{ppK:Akaishi}
are of comparable quality. 

The importance of the $T_N=0$ component of the wave function is underlined by the comparison
between ``Model I'' with ``Model II''. Although this admixture is only about 6 \% as shown in 
the last row [$P(T_N=0)$] in Table \ref{tab:Bench}, switching it off reduces the binding energy by more
than 20 \%.  The matrix elements of the $\bar{K}N$ interaction 
taken between normalized states $| \Phi_+ \rangle$ and $| \Phi_- \rangle$ are as follows: 
\begin{eqnarray}
\langle \Phi_+ | \; \hat{V}_{\bar{K}N} \; | \Phi_+ \rangle 
& = &  \frac{3}{4} \, v_{\bar{K}N}^{I=0} \; + \; \frac{1}{4} \, v_{\bar{K}N}^{I=1} ,\\
\langle \Phi_- | \; \hat{V}_{\bar{K}N} \; | \Phi_- \rangle  
& = &  \frac{1}{4} \, v_{\bar{K}N}^{I=0} \; + \; \frac{3}{4} \, v_{\bar{K}N}^{I=1} ,\\
\langle \Phi_+ | \; \hat{V}_{\bar{K}N} \; | \Phi_- \rangle
& = &  \frac{\sqrt{3}}{4} \, \left( v_{\bar{K}N}^{I=0} \; - \; v_{\bar{K}N}^{I=1} \right) .\label{Mat:mix}
\end{eqnarray}
The mixture implied by Eq.~(\ref{Mat:mix}) increases the $ppK^-$
binding energy. For the AY potential the values of these three matrix elements are
$-173.3$ MeV, $-112.3$ MeV 
and $-53.0$ MeV, in this order. Taking the mixing ratio between $| \Phi_+ \rangle$ and 
$| \Phi_- \rangle$ into account, the actual contributions from each matrix element to the $\bar{K}N$ potential energy are $-163.1$ MeV, $-6.6$ MeV and $-24.9$ MeV, respectively, 
and so the coupling matrix element (\ref{Mat:mix}) is found to be attractive and non-negligible.

In concluding this Appendix we note again that the overall attraction produced by the 
$\hat{V}_{\bar{K}N}$ based on chiral SU(3) dynamics, and used in the present work, is considerably weaker in comparison and leads to a $ppK^-$ binding energy less than half of that found with the simple AY potential.

\section{Dependence on the $NN$ potential\label{Sec:Minnesota}}

\begin{figure}[t]
    \centering
    \includegraphics[width=0.45\textwidth,clip]{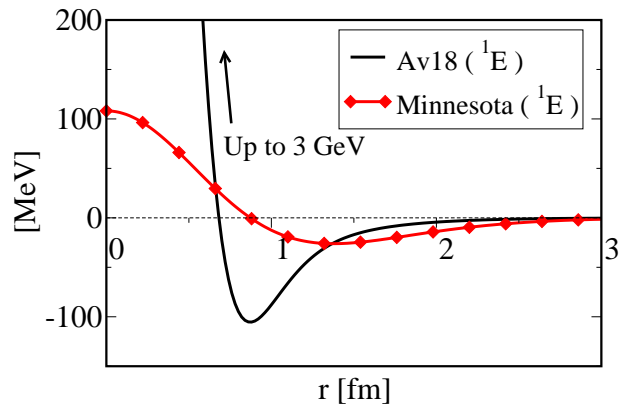}
    \caption{\label{fig:NNpot_Minn} 
    (Color online) Comparison of Minnesota potential with 
    Av18 potential. $^1E$ channel of both potentials are shown. 
    Minnesota potential is depicted with a solid line with diamond. 
    Av18 potential is depicted with a solid line. 
}
\end{figure}%

Test calculations have been performed replacing the Av18 by a 
soft-core $NN$ potential, the Minnesota potential~\cite{NN:Minnesota}
(see Fig.~\ref{fig:NNpot_Minn}).
This potential reproduces low-energy $NN$ data (scattering lengths, 
effective ranges, and deuteron properties) just like the more 
realistic Av18 interaction. In the actual computations, we have fixed 
a parameter of the Minnesota potential ($u=1$) so that it 
corresponds to a Serber-type potential. 

\begin{table}[t]
\caption{\label{tab:Minnesota} 
Dependence of $ppK^-$ results on the $NN$ potential.
The fist and second lines indicate the type of $\bar{K}N$ and $NN$ potential, 
respectively. ``Minnesota'' in the second line 
refers to the soft-core potential of Ref.~\cite{NN:Minnesota}.
$E_{nuc}$ is the total energy of nuclear part, namely $E_{nuc}=T_{nuc}+V(NN)$. 
Other quantities shown in this table are the same as those in Table \ref{tab:Bench}. 
All results are obtained with Gaussian expansions $N_N=N_K=5$ 
in Eqs. (\ref{NNcorr}) and (\ref{KNcorr}). 
}
\begin{ruledtabular}
\begin{tabular}{l|rr|rr}
$\bar{K}N$     & HNJH    &           & AY      &           \\
$NN$           & Av18    & Minnesota & Av18    & Minnesota \\
\hline
               &         &           &         &           \\
$B(ppK^-)$     & 16.9    &   17.0    &  49.0   & 50.4      \\
$\Gamma$       & 47.0    &   49.4    &  60.1   & 64.7      \\
               &         &           &         &           \\
$B_K$          & 38.9    &   42.1    &  78.1   & 84.2      \\
$E_{nuc}$      & 22.0    &   25.1    &  29.1   & 33.8      \\
$T_{nuc}$      & 38.1    &   32.9    &  49.5   & 41.3      \\
               &         &           &         &           \\
$E_{kin}$      & 129.5   &   131.3   & 160.6   & 160.0     \\
$V(NN)$        & $-$16.2 & $-$7.8    & $-$20.5 & $-$7.5    \\
$V(\bar{K}N)$  & $-$130.4& $-$140.5  & $-$189.2& $-$202.9  \\
               &         &           &         &           \\
$R_{NN}$       & 2.21    & 2.15      & 1.82    & 1.80      \\
$R_{\bar{K}N}$ & 1.97    & 1.93      & 1.56    & 1.53      \\
               &         &           &         &           \\
$P(T_N=0)$     & 3.8 \%  & 4.5 \%    & 4.4 \%  & 5.1 \%    \\
               &         &           &         &           
\end{tabular}
\end{ruledtabular}
\end{table}

\begin{figure}[b]
    \centering
    \includegraphics[width=0.45\textwidth,clip]{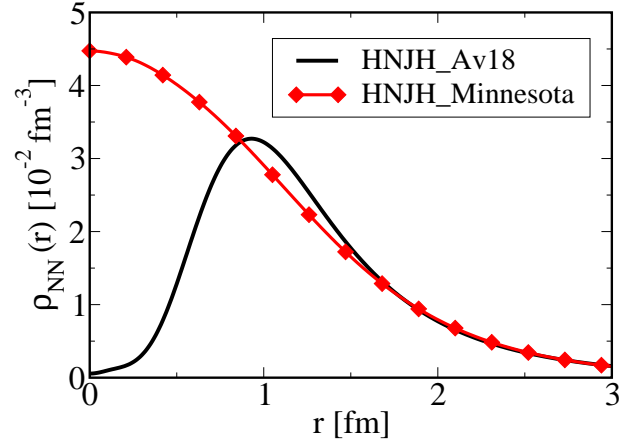}
    \caption{\label{fig:NNdens_Minn} 
    (Color online) $NN$ density for Av18 potential (solid line) and 
    Minnesota potential (solid line with diamond). 
    A chiral-based $\bar{K}N$ potential (HNJH) is used. 
}
\end{figure}%

Results of $ppK^-$ calculations are summarized in Table~\ref{tab:Minnesota}
for the chiral $\bar{K}N$ potential (HNJH) and for the phenomenological AY 
potential. The sensitivity to details of the $NN$ interaction turns 
out to be marginal for the weakly bound $ppK^-$ system and slightly 
more pronounced but still weak for the more strongly bound AY case. 
The qualitative difference between strong and soft short-range 
repulsive core becomes apparent, however, when examining the $NN$ 
density distribution within the $\bar{K}NN$ clusters 
(see Fig.~\ref{fig:NNdens_Minn}).

\end{document}